\documentclass[fleqn,usenatbib]{mnras}

\usepackage{newtxtext,newtxmath}

\usepackage[T1]{fontenc}
\usepackage{array}
\usepackage{booktabs}
\usepackage{multirow}
\DeclareRobustCommand{\VAN}[3]{#2}
\let\VANthebibliography\thebibliography
\def\thebibliography{\DeclareRobustCommand{\VAN}[3]{##3}\VANthebibliography}

\usepackage{graphicx}	
\usepackage{amsmath}	

\newcommand{\rp}{R_{\mathrm{p}}}
\newcommand{\muv}{M_{\mathrm{UV}}}
\newcommand{\meanMuv}{M_{\mathrm{UV}}^{\star}}

\newcommand{\Rp}{$R_{\mathrm{p}}\ $}
\newcommand{\Muv}{$M_{\mathrm{UV}}\ $}
\newcommand{\MeanMuv}{$M_{\mathrm{UV}}^{\star}\ $}

\newcommand{\hi}{H\,\textsc{i}}

\newcommand{\hei}{He\,\textsc{i}}
\newcommand{\heii}{He\,\textsc{ii}}
\newcommand{\Lya}{Lyman $\alpha \ $}

\title{Modeling Quasar Proximity Zones in a Realistic Cosmological Environment with a Self-consistent Light Curve}

\author[Zhou et al. 2023]{
Yihao Zhou,$^{1}$\thanks{E-mail: yihaoz@andrew.cmu.edu}
Huanqing Chen,$^{2}$
Tiziana Di Matteo,$^{1}$
Yueying Ni,$^{3}$ 
Rupert A.C. Croft, $^{1}$ and
Simeon Bird$^{4}$
\\
$^{1}$McWillams Center for Cosmology, Department of Physics, Carnegie Mellon University, Pittsburgh, PA 15213, USA\\
$^{2}$Canadian Institute for Theoretical Astrophysics, University of Toronto,60 St George St, Toronto, ON M5R 2M8, Canada\\
$^{3}$Harvard-Smithsonian Center for Astrophysics, 60 Garden Street, Cambridge, MA 02138, USA\\
$^{4}$Department of Physics \& Astronomy, University of California, Riverside, 900 University Ave., Riverside, CA 92521, USA
}

\date{Accepted XXX. Received YYY; in original form ZZZ}

\pubyear{2023}

\begin{document}
\label{firstpage}
\pagerange{\pageref{firstpage}--\pageref{lastpage}}
\maketitle

\begin{abstract}

We study quasar proximity zones in a simulation that includes a self-consistent quasar formation model and realistic IGM environments.
The quasar host halo is $10^{13}\ M_{\mathrm{\odot}}$ at $z=6$, more massive than typical halos studied in previous work.
Between $6<z<7.5$, the quasar luminosity varies rapidly, with a mean magnitude of $M_{\mathrm{UV,mean}}=-24.8$ and the fluctuation reaching up to two orders of magnitude.
Using this light curve to post-process the dense environment around the quasar, we find that the proximity zone size ($R_{\mathrm{p}}$) ranges between $0.5$-5 pMpc. 
We show that the light curve variability causes a similar degree of scatter in \( R_{\mathrm{p}} \) as does the density fluctuation, both of which result in a standard deviation of \( \sim 0.3\ \mathrm{pMpc} \).
The $R_{\mathrm{p}}$ traces the light curve fluctuations closely but with a time delay of $\sim 10^4\ \mathrm{yr}$, breaking the correspondence between the $R_{\mathrm{p}}$ and the contemporaneous $M_{\mathrm{UV}}$. This also indicates that we can only infer  quasar activity within the past $\sim 10^4$ years instead of the integrated lifetime from $R_{\mathrm{p}}$ in the later part of cosmic reionization. 
Compared with the variable light curve, a constant light curve underestimates the $R_{\mathrm{p}}$ by 13\% at the dim end ($M_{\mathrm{UV}}\sim -23.5$), and overestimates the $R_{\mathrm{p}}$ by 30\% at the bright end ($M_{\mathrm{UV}}\sim -26$). 
By calculating the $R_{\mathrm{p}}$ generated by a number of quasars, we show that variable light curves predict a wider $R_{\mathrm{p}}$ distribution than lightbulb models, and readily explain the extremely small $R_{\mathrm{p}}$ values that have been observed.

\end{abstract}




\section{Introduction}
\label{sec:intro}
A bright quasar at high redshift usually creates a large region, commonly referred to as a `quasar proximity zone', where the ionizing radiation contributed from the quasar significantly exceeds the cosmic ionizing background. Within a quasar proximity zone, the hydrogen neutral fraction is considerably lower than typical regions in the Universe. At $z>6$, these are the only regions where we can observe non-zero \Lya transmitted flux \citep{Bajtlik1988, Cen2000, Wyithe2005, Bolton2007_NR_RT_model, Bolton2007, Lidz2007}. As a result, quasar proximity zones are unique windows for probing the distant universe. 

One key observational measurement related to a quasar proximity zone is its size, which is traditionally defined in quasar spectra as the distance from the systematic \Lya line center to the first point where the transmitted flux drops below 10\% of the continuum level after being smoothed by a 20\AA\ top-hat kernel \citep{Fan2006_qso_obs, Carilli2010_qso_ob, Eilers2017_qso_obs, Eilers2020_qso_obs, Mazzucchelli2017, Ishimoto2020_qso_ob}. \cite{Fan2006_qso_obs} compiled the first large sample of $z \gtrsim 6$ quasar spectra 
and measured a proximity zone size $R_{\mathrm{p}} \sim 8\ \mathrm{Mpc}$ at $z = 6$ for quasars of magnitude $M_{\mathrm{1450}}=27$. 
\cite{Carilli2010_qso_ob} analyzed the proximity zone sizes of 27 quasars with more accurate redshift measurements. In the last decade, the number of high redshift quasar spectra with well-measured \Lya proximity zone sizes has grown considerably \citep{Reed2017_highz_qso, Banados2018_qso_75, Matsuoka2019_highz_qso, Wang2019_highz_qso}.
For example, \cite{Eilers2017_qso_obs} studied quasar proximity zones in the redshift range $5.77 \leqslant z \leqslant 6.54$ with a homogeneous analysis of 34 medium resolution spectra, and \cite{Ishimoto2020_qso_ob} presented measurements of the proximity zone size for 11 low-luminosity ($-26.16 \leqslant M_{\mathrm{1450}} \leqslant -22.83$) quasars at $z\sim 6$. An unexpected result from these observations obtained in recent years is that some quasars display very small proximity zones, such as the \Rp$\sim 0.37$ Mpc measured in \cite{Eiler2021_young_qso}. 
One possible explanation for several small $R_{\mathrm{p}}$ seen in $z\gtrsim 7$ quasar spectra is that a large amount of hydrogen at $z\sim 7$ is still neutral; e.g.,  \citealt{Miralda-Escude1998,Mortlock2011_qso_70, Bolton2011_qso_7_IGM, Bosman2015_qso_7_IGM, Banados2018_qso_75, Davies2018,Wang2020_dampingwing, Yang2020_dampingwing, Bosman2015_11200641, Greig2017_11200641, Greig2022_dampingwing}). However, the population of small $R_\mathrm{p}$ quasars at $z\sim 6$ remains perplexing.

The sizes of quasar proximity zones could provide valuable insights into quasar activity and cosmic reionization, an epoch when the IGM transitioned from a mostly neutral state into a mostly ionized state. Several physically motivated (semi-)analytic models of proximity zones have been proposed, which lead to scaling relations between $R_{\mathrm{p}}$ and the intrinsic properties of quasars (e.g., the luminosity and the lifetime) as well as the surrounding IGM.
In a nearly neutral universe, the proximity zone size is closely related to the size of the quasar ionized bubble, which is sensitive to the ionization fraction of the local IGM and the total number of ionizing photons the quasar emits \citep{Cen2000, Haiman2001}:
\begin{equation}
\label{equ:R_ion_one_third}
    R_{\mathrm{ion}} = \left(\frac{3\dot{N}t_{\mathrm{q}}}{4\pi\,n_{\mathrm{H}}\,x_{\mathrm{\hi}}} \right)^{1/3},
\end{equation}
where $\dot{N}$ is the emitted ionizing photon rate, $t_{\mathrm{q}}$ is the quasar lifetime, $n_{\mathrm{H}}$ is the hydrogen number density and $x_{\mathrm{\hi}}$ is the neutral hydrogen fraction. 
On the other hand, if the quasar is embedded in an already ionized IGM,
\citet{Bolton2007_NR_RT_model} showed that the proximity zone size quickly reaches a maximum 
$R_{\mathrm{p}}^{\mathrm{max}}$ which is independent on the neutral fraction:
\begin{equation}
\label{equ:ana_Rp_BH2007}
\begin{aligned}    
    R_{\mathrm{p}}^{\mathrm{max}} &= \frac{3.14}{\Delta_{\mathrm{lim}}}\ \left( \frac{\dot{N}}{2\times 10^{57}\ \mathrm{s}^{-1}}\right)^{1/2}\ 
     \left(\frac{T}{2\times10^{4} \ \mathrm{K}}\right)^{0.35}\\  
     &\times\left(\frac{\tau_{\mathrm{lim}}}{2.3}\right)^{1/2}\,  \left(\frac{\alpha^{-1}\left(\alpha+3\right)}{3}\right)^{-1/2}
     \left( \frac{1+z}{7}\right)^{-9/4}
     \ \ \ \mathrm{pMpc}.
    \end{aligned}    
\end{equation}
Here $\tau_{\mathrm{lim}}=-\ln(F_{\mathrm{lim}})$ is the limiting \Lya optical depth (typically $F_{\mathrm{lim}}$ is set to be $0.1$) and $\alpha$ denotes the spectral index of the quasar power-law spectrum. 
Recently, \cite{Davies2020_NR_RT_model} developed a semi-analytic model to trace the time-dependent evolution of the mean proximity zone size, considering both the photoionization and photoheating effects.

Although (semi-)analytical models delineate the typical trend in proximity zone size evolution, it is difficult for them to describe the inhomogeneity of the IGM, and the complex dependence of $R_{\mathrm{p}}$ on quasar emission history. As a result, many works combine hydrodynamic simulations and post-processing with radiative transfer (RT) to investigate the formation and the properties of proximity zones. For example, 
\cite{Lidz2007} calculated the mock \Lya flux in quasar proximity zones from simulations and showed that the absorption spectra are sensitive to the level of small-scale structure in the IGM. They suggested that the proximity zone measurements are compatible with the fact that $z\sim 6$ quasars reside in a pre-ionized IGM.
\cite{Keating2015} used simulations that included a wide range of host halo masses ($10^{10-12.5} \ M_{\odot}$) and discovered that $R_{\mathrm{p}}$ depends weakly on the mass of the host halo, but strongly on direction, leading to a large $R_{\mathrm{p}}$ scatter across lines of sight.
\cite{Chen2021_qso_prixmity_zone} found that about $1\%\sim 2\%$ of old quasars exhibit extraordinarily small proximity zone sizes. Their results are based on the Cosmic Reionization On Computers (CROC) simulations, which can resolve Lyman Limit Systems (LLSs) because of its relatively high spatial resolution.
These small proximity zones are predominantly due to the occurrence of a damped Ly$\alpha$ absorber (DLA) or an LLS along the line of sight. They can be distinguished from the small proximity zones of young quasars because of the metal contamination.

Most of the modeling of proximity zones using cosmological simulations post-processed with RT assumes a `lightbulb' model for quasar light curves. In such a model, the quasar switches on abruptly and then maintains the same luminosity. 
However, realistic quasar light curves are much more complex. They may instead undergo short periods of flickering, with the fluctuation magnitude as well as the timescale varying continuously \citep{Novak2011_lc_vary, Gabor2013_qso_lc, King2015_qso_lc, Schawinski2015_qso_lc, Oppenheimer2018_qso_lc, Shen2021_qso_lc}. Extreme variability is observed in some quasars with  magnitude changes of $\sim 1\ \mathrm{mag}$ within $\sim 15\ \mathrm{yr}$ \citep{Rumbaugh2018_qso_vary}.
Furthermore, lightbulb models can hardly explain the small proximity zones observed around luminous quasars at $z\sim 6$. Although 
\citet{Eilers2017_qso_obs, Eiler2021_young_qso} offer the suggestion that these small proximity zones are from young quasars that just turned on for $t_\mathrm{q}=10^4$ yr,
these estimated lifetime values are significantly smaller than the timescales required by current supermassive black hole (SMBH) formation models \citep{Volonteri2010_SMBH_model, Li2012_SMBH_acc, Madau2014, 
Volonteri2015, Smith2017,
Weinberger2018, Regan2019,
Inayoshi2020}. 

To model quasar variability, several works have built toy models for flickering light curves, most of which have fixed duty cycle $f_{\mathrm{duty}}$ or episodic lifetime $t_{\mathrm{ep}}$ \citep{ Soltinsky2023_qso_lifetime, Satyavolu2023_qso_growth}, and predicted a very different proximity zone evolution. \cite{Davies2020_NR_RT_model} applied a light curve consisting of different time-scale variations and computed $R_{\mathrm{p}}$ based on their time-dependent semi-analytical model. They found that a variable light curve yielded a $R_{\mathrm{p}}$ distribution strongly skewed towards small values compared to a typical lightbulb model. 

In this work, we go a step further and study a more realistic quasar light curve. We utilize data from a state-of-the-art cosmological simulation that incorporates subgrid models of galaxy and black hole formation, and extract the light curve from the simulated active galactic nuclei (AGNs).
We model the accretion onto AGNs in the same fashion as the \textsc{BLUETIDES} \citep{Feng2016_Bluetides} and \textit{MassiveBlack I \& II} \citep{dimatteo2012, Khandai2015_massiveBH} simulations. 
They achieve close agreement with observational measurements of the relationships between SMBHs and their host galaxies \citep{DeGraf2015_simu_obs_test, Ding2020_simu_obs_test, Ding_2022_simu_obs_test}, the  observed quasar luminosity functions (QLF), and the correlation length $r_{0}$ of AGN clustering \citep{Khandai2015_massiveBH}. 

The rest of this paper is organized as follows. In Section~\ref{sec:method}, we briefly summarize the basic features of the cosmological simulations used in this work and
then illustrate the method used to compute proximity zone size evolution with RT. In Section~\ref{section: results}, we present our results, and
discuss our key results in Section~\ref{section:discuss}. We conclude with Section~\ref{section:conclusion}.

\section{Methodology}
\label{sec:method}

\subsection{Constrained realization simulation}
\label{sec:simulation}

\begin{figure*}
	\includegraphics[width=1\textwidth]{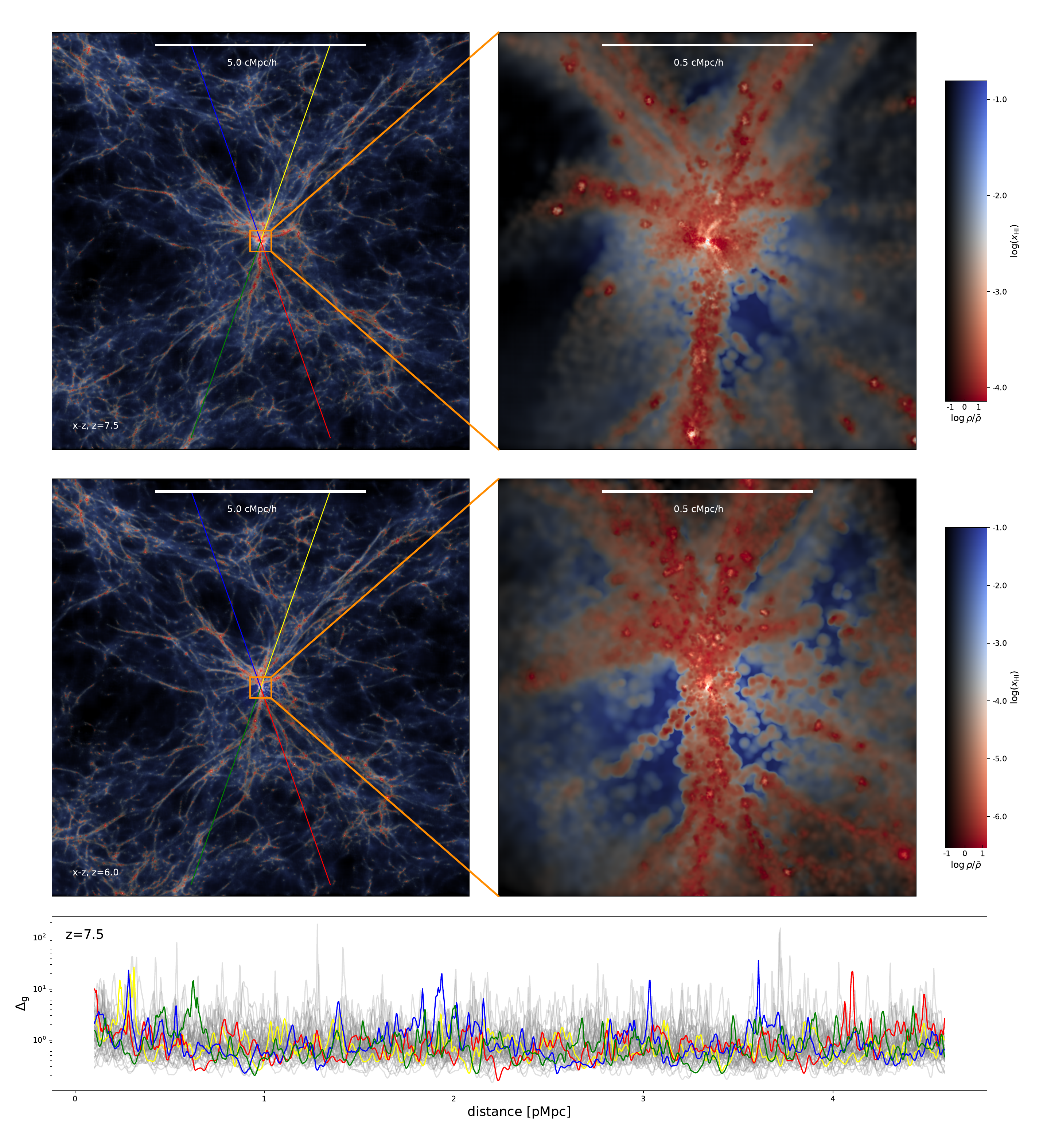}
    \caption{Illustration of the host environment of the quasar prior to the RT post-processing. The first two rows are the snapshots of the gas density fields, with the quasar positioned at the center of each panel, at $z=7.5$ (upper) and $z=6.0$ (middle). 
    For regions with high density, the color hue is set by the hydrogen neutral fraction $x_{\mathrm{\hi}}$. The sidebar  transitions from red to blue, representing the progression from ionized to neutral states. 
    The left two panels display cubical boxes with a comoving side length of $10\ h^{-1}\mathrm{Mpc}$, and the two right panels depict the zoomed-in central regions of diameter $ 1\ h^{-1}\mathrm{Mpc}$.
    The blue, yellow, green, and red straight lines in the left panels show the directions of 4 lines of sight used in this work. In the bottom panel, we show the gas density contrast $\Delta_{\mathrm{g}}$ ($=\rho/\bar{\rho}$) along all 48 directions at $z=7.5$ in grey, with the four examples in the upper panels highlighted in blue, yellow, green, and red, respectively. 
    }
    \label{fig:gas_map}
\end{figure*}

\begin{figure}
	\includegraphics[width=\columnwidth]{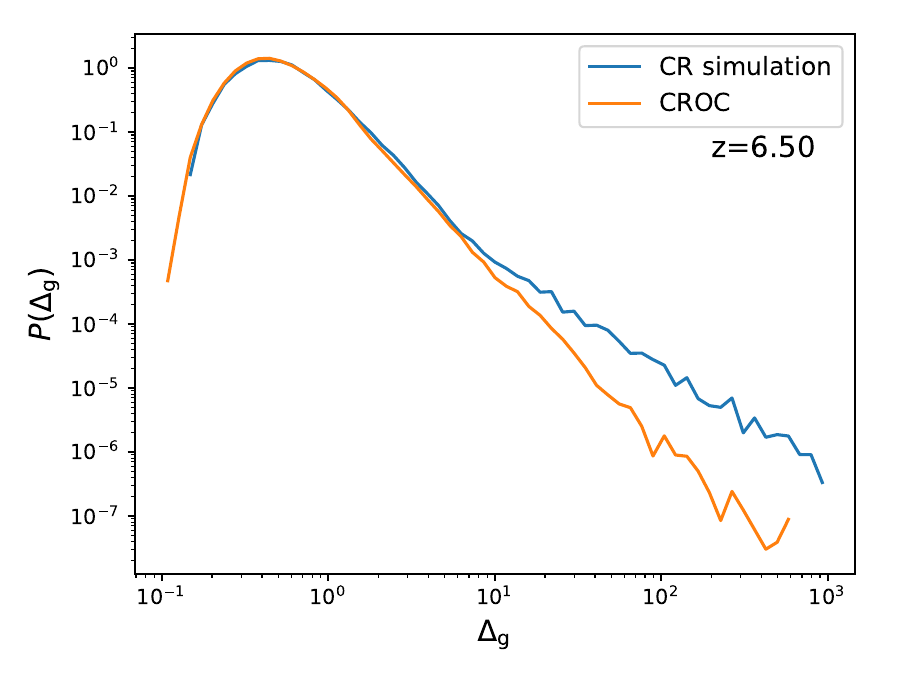}
    \caption{The gas density contrast ($\Delta_{\mathrm{g}}$) PDF for the lines of sight from the CR simulation used in this work (blue), and those from the CROC simulation (orange) used in \protect\cite{Chen2021_qso_prixmity_zone}. Each pixel is $4\ \mathrm{pkpc}$ in length, and all the data are drawn from $0.1 \sim 2\ \mathrm{pMpc}$ regions from the quasar in the $z=6.5$ snapshot.}
    \label{fig:los_PDF}
\end{figure}

To simulate the proximity zones around a long-lived luminous quasar, 
we use one of the cosmological simulations with constrained realizations (CR) described in \cite{Ni2021_gaussianCR}. 
Early quasars are extremely rare, and generally, only large volume cosmological simulations (with box size $\gtrsim 500 \ \mathrm{cMpc}$) can capture them. However, an alternative method is to use the CR technique to model these rare objects.
The CR technique \citep{Hoffman1991_CR, vandeWeygaert1996_CR_para} imposes various user-specified constraints on the Gaussian random field of the initial conditions (ICs).
Within a relatively small box, the CR technique allows us to model a rare high-density peak in the primordial density field that will form a massive halo hosting a bright quasar at high redshift ($z\sim 7$). 
Among the CR simulations carried out in \cite{Ni2022_astrid}, we select the one that contains the quasar with the largest instantaneous bolometric luminosity, since we focus on the brightest quasars in this study.
The quasar's black hole mass is  $M_{\mathrm{BH}}\sim 10^{9}\ M_{\odot}$ at $z=6$.

The CR simulation has a box size of $20\,h^{-1}\ \mathrm{cMpc}$ per side. The ICs are generated using the most recent implementation of \texttt{gaussianCR}\footnote{\url{https://github.com/yueyingn/gaussianCR}}, and the simulation is  run  to $z=6$ with
the massively parallel cosmological smoothed-particle hydrodynamics (SPH) simulation software \texttt{MP-Gadget}  \citep{Feng2018_MPGadget}.
The peak height of our quasar-hosting halo is $5\,\sigma_{0}$ with a peculiar velocity $\boldsymbol{v}=\boldsymbol{0}\,
\text{km}\,\text{s}^{-1}$, where $\sigma_{0}$ is the density variance after convolution with a Gaussian kernel of  size  $R_{\mathrm{G}}=1\ h^{-1}\, \text{Mpc}$.
For full details of the underlying formalism of the CR simulation, we refer the reader to \cite{vandeWeygaert1996_CR_para, Ni2021_gaussianCR}.

The box contains $2\times352^{3}$ particles, with a mass resolution $M_{\mathrm{DM}} = 1.2\times 10^{7}M_{\odot}\,h^{-1}$ for the dark matter particles and $M_{\mathrm{gas}} = 2.4\times10^{7}M_{\odot}\,h^{-1}$ for the gas particles in the ICs.
Star particles have a mass of $M_{*} = 6\times 10^{5}M_{\odot}\,h^{-1}$. The gravitational smoothing length is $1.8\,h^{-1} \text{ckpc}$ for both dark matter and gas particles. The cosmological model is consistent with the nine-year \emph{Wilkinson Microwave Anisotropy Probe} (WMAP) data \citep{Hinshaw_2013_WMAP} ($\Omega_{0} = 0.2814$, $\Omega_{\Lambda} = 0.7186$, $\Omega_{\mathrm{b}} = 0.0464$, $\sigma_{8} = 0.82$, $h=0.697$, $n_{s}=0.971$).

\subsection{Physics: hydrodynamics and sub-grid modelling}
\label{section:simu_phy}

Most of the model physics implemented in the CR simulation is the same as in \textsc{BLUETIDES} \citep{Feng2016_Bluetides}. 
The gravity is solved with the treePM approach.
The pressure-entropy formulation of SPH is adopted to solve the Euler equations \citep{Read2010_pSPH, Hopkins2013_pSPH}. The density estimator uses a quintic kernel to reduce the noise in the SPH density and gradient estimation \citep{Liu_SPH_kernel}. 
A range of sub-grid models are applied to simulate galaxy and black hole formation and associated feedback processes. 
Radiative cooling from both primordial gas \citep{Katz1996_gas_cooling_radia} and metals \citep{Vogelsberger2014_gas_metal_cool} is considered. 
Star formation is implemented based on the multiphase star formation model \citep{Springel2003_SF_model},  but incorporating several effects described in
\cite{Vogelsberger2013_SF_model_modification}. The formation of molecular hydrogen is computed according to the prescription of \cite{Krumholz2011_sfr}, and its effect on star formation at low metallicities is considered.  Type II supernova wind feedback is included, using the same model as in the Illustris simulation \citep{Nelson2015_illustris, Okamoto2010_SNII_feedback}. Their wind speeds are assumed to be proportional to the local one-dimensional dark matter velocity dispersion $\sigma_{\mathrm{DM}}$: $v_{\mathrm{w}} = \kappa_{\mathrm{w}}\,\sigma_{\mathrm{DM}}$, where $v_{\mathrm{w}}$ is the wind speed, and the dimensionless parameter $\kappa_{\mathrm{w}}=3.7$ \citep{Vogelsberger2013_SF_model_modification}.

Black hole growth and AGN feedback are modeled in the same way as in the \textit{MassiveBlack I \& II} simulations, based on the black hole sub-grid model developed in \cite{Springel2005_bh_model} and \cite{DiMatteo2005_BH_model}.
Black holes are seeded with an initial seed mass of $M_{\mathrm{seed}} = 5 \times 10^{5} M_{\odot}\,h^{-1}$ in halos with a mass larger than $5\times 10^{10} M_{\odot}\,h^{-1}$. 
Note that our choice of seed mass is close to that expected from direct collapse scenarios (e.g., \citealt{Latif2013_dc_bh_seed, Schleicher2013_dc_bh_seed, Ferrara2014_dc_bh_seed}), but our seeding scheme makes no direct assumption for the black hole seed formation mechanism. 

The gas accretion rate of the black hole is given by the Bondi-Hoyle rate \citep{Bondi_Hoyle_1944_accretion_rate}: 
\begin{equation}    
\label{equ:Mdot}
\dot{M}_{\mathrm{B}}=4\pi\,G^{2}\,M_{\mathrm{BH}}^{2}\,\rho_{\mathrm{BH}}\left(c_{\mathrm{s}}^{2}+v^{2}_{\mathrm{vel}}\right)^{-3/2}, 
\end{equation}
where $c_{\mathrm{s}}$ is the local sound speed, $\rho_{\mathrm{BH}}$ is the gas density around the quasar, and $v_{\mathrm{vel}}$ is the velocity of the black hole relative to the surrounding gas. 
Super-Eddington accretion is allowed with an upper limit of twice the Eddington accretion rate $\dot{M}_{\mathrm{Edd}}$. Therefore the black hole accretion rate $\dot{M}_{\mathrm{BH}}$ is determined by $\dot{M}_{\mathrm{BH}} = \min\left(\dot{M}_{\mathrm{B}}, 2\dot{M}_{\mathrm{Edd}}\right)$. With a radiative efficiency $\eta=0.1$ \citep{Shakura1973_BH}, the black hole radiates with a bolometric luminosity $L_{\mathrm{bol}}$ proportional to the accretion rate: $L_{\mathrm{bol}}=\eta\,\dot{M}_{\mathrm{BH}} c^{2}$. 
Five percent of the radiated energy is thermally coupled to the gas residing within twice the radius of the SPH smoothing kernel of the black hole particle, which is typically about $1\% \sim 3\%$ of the virial radius of the halo. 

The patchy reionization model \citep{Battaglia2013_patchy_reion} is not included in the CR simulation because of the small box, instead the reionization is assumed to occur instantaneously.
The applied global ionization history is consistent with that in \textsc{BLUETIDES}, and reionization is almost completed at redshift $z=8$ (Fig. 2 in \cite{Feng2016_Bluetides}). Consequently, the gas in the snapshots we use herein  ($z \leqslant 7.5$) is originally highly ionized.

\begin{figure*}
	\includegraphics[width=1\textwidth]{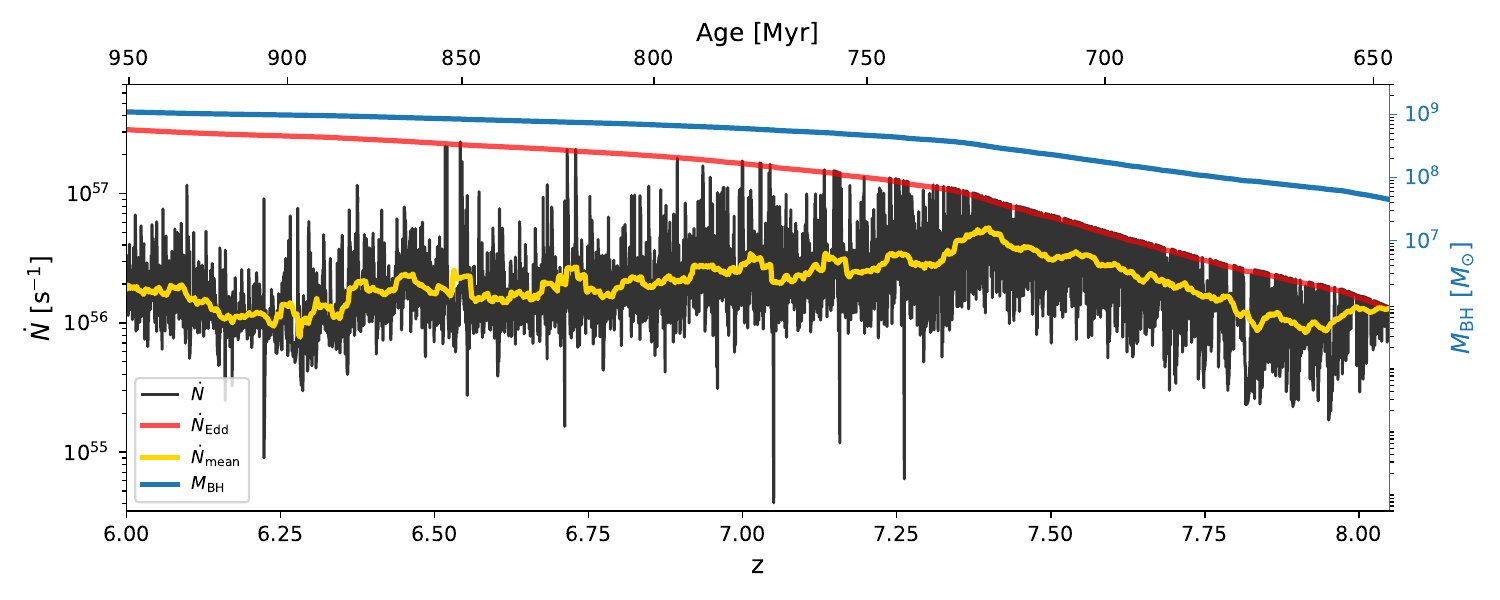}
    \caption{The mass evolution (blue curve; using right y-axis) and the ionizing photon number emitted  per second by the accretion disk (black curve; left y-axis) of the brightest quasar in the simulation. The red solid line represents the photon number rate corresponding to twice the Eddington limit (upper limit set by our simulation). The yellow curve denotes the averaged photon number rate $\dot{N}_{\mathrm{mean}}$ computed over a time kernel of $5\ \mathrm{Myr}$.
    The upper axis indicates the age of the Universe at a given redshift.}
    \label{fig:qso_LC}
\end{figure*}

\begin{figure*}
	\includegraphics[width=1\textwidth]{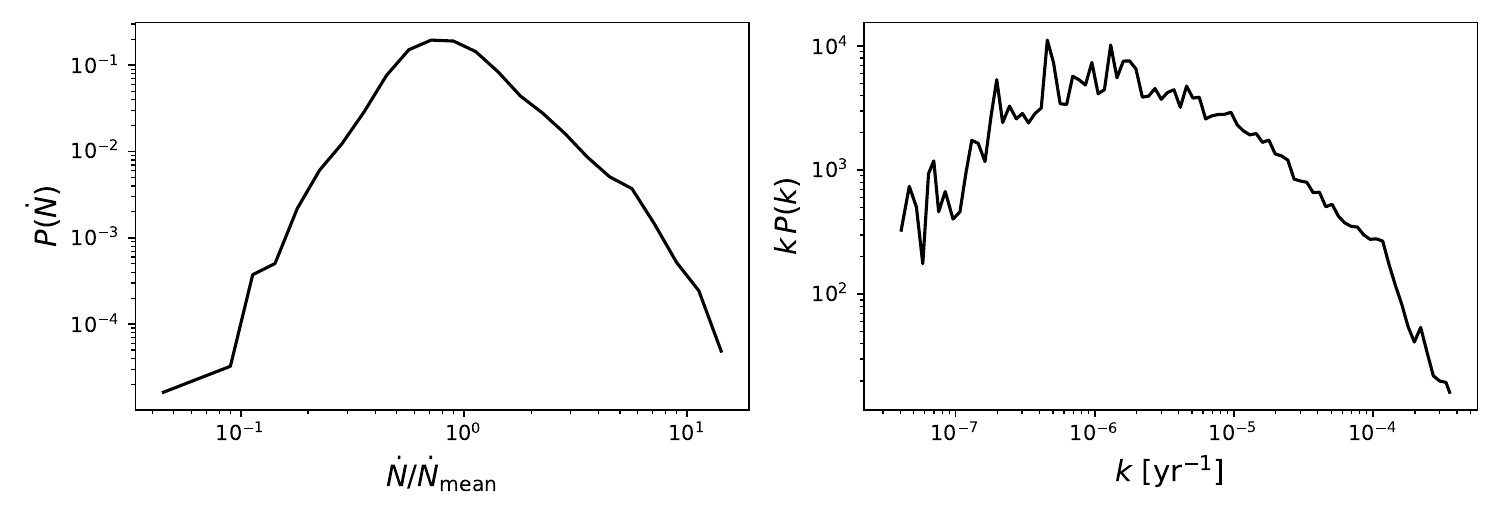}
    \caption{The statistics for the quasar light curves with $6\leqslant z \leqslant 7.5$. 
    Left panel: the PDF of the ionizing photon rate contrast $\dot{N}/\dot{N}_{\mathrm{mean}}$, where $\dot{N}_{\mathrm{mean}}$ is the mean of the ionizing photon number rate $\dot{N}$ emitted by the quasar.
    Right panel: the dimensionless power spectrum $kP(k)$ for the time evolution of $\dot{N}/\dot{N}_{\mathrm{mean}}$.}
    \label{fig:qso_LC_stat}
\end{figure*}

\subsection{Line of sight gas densities}
\label{section:los}
The projected gas density fields around  
the quasar at $z=7.5$ and $z=6.0$ are shown in the first two rows of Fig.~\ref{fig:gas_map}, with the quasar located at the center of each panel. 
Regions with high density are color-coded by the hydrogen ionization fraction $x_{\mathrm{\hi}}$, from red to blue indicating ionized ($x_{\mathrm{\hi}}\lesssim 10^{-4}$) to neutral ($x_{\mathrm{\hi}}\sim 0.1$), as shown by the color bars.
The left two panels display the regions  
$10\ h^{-1}\mathrm{Mpc}$ in width centered on the quasar, while the right two panels depict the central zoomed-in regions of width $1 \ h^{-1}\mathrm{Mpc}$.

We use \textsc{HEALPY}\footnote{\url{https://github.com/healpy/healpy}} to cast 48 evenly spaced lines of sight,
starting from the position of the quasar, and employ the SPH formalism to calculate the gas properties $f(\boldsymbol{x})$ (e.g., density, velocity, ionization fraction)
at position $\boldsymbol{x}$ on the line of sight \citep{Liu_SPH_kernel}:
\begin{equation}
    \left<f(\boldsymbol{x})\right> = \sum_{j=1}^{N}
    \frac{m_{j}}{\rho_{j}} 
    f\left(\boldsymbol{x}_{j}\right)\,W\left(\boldsymbol{x}-\boldsymbol{x}_{j},q\right),
\end{equation}
where $\sum_{j}$ is the sum over all the neighboring gas particles within the smoothing length $q$,
and $W$ is the quintic density kernel.
$m_{j}$, $\rho_{j}$, and $\boldsymbol{x}_{j}$ are the mass, density, and position of each particle, respectively.
Taking advantage of the periodic boundary conditions of the simulation box, we extend each line of sight to a length of $40\,h^{-1}\text{Mpc}$. The sightlines are drawn significant off-axes (not parallel to the $x$, $y$ or $z$ axes), ensuring that none of them travel through the massive halo again.
The spatial resolution is set to be $30$ comoving kpc, equivalent to $\sim 4$ kpc in proper units.
We indicate the directions of 4 lines of sight in the left panels in Fig.~\ref{fig:gas_map} (blue, yellow, green, and red straight lines).
In the bottom panel, we show the gas density contrast $\Delta_{\mathrm{g}}(=\rho/\bar{\rho}$; where $\bar{\rho}$ is the mean gas density of the Universe) 
at $z=7.5$ for all the 48 lines of sight. The four examples shown in the upper panels are also included, represented by correspondingly colored lines.  
These $\Delta_{\mathrm{g}}$ profiles enable us to see that the density peaks in different lines of sight are located at different radii, and that
the density fluctuations along each direction span about two orders of magnitude.

One of the unique properties of our simulation is the constrained initial conditions. It is thus useful to compare our sightlines with one of the simulations studied in \cite{Chen2021_qso_prixmity_zone}, which models a more commonly occurring  type of region without constrained conditions. \cite{Chen2021_qso_prixmity_zone} studied sightlines drawn from the B40E CROC simulation, where the box size is $40\,h^{-1}\ \mathrm{cMpc}$ on each side. The CROC simulation is run with the Adaptive Refinement Tree code \citep{kravtsov1999, kravtsov2002, rudd2008}, with a base resolution of $39\,h^{-1} \mathrm{ckpc}$ and a peak resolution of 100 pc. All the sightlines are drawn from halos with dark matter mass larger than $1.5\times10^{11} \ M_\odot$. Another major difference is that the CROC simulation models reionization by star particles self-consistently, resulting in a volume-weighted neutral fraction $\left<x_{\mathrm{\hi}}\right>_{\rm v}=0.13$ at $z=7.33$ and $\left<x_{\mathrm{\hi}}\right>_{\rm v} < 6.7\,\times10^{-4}$ after $z=6.7$.

In Figure \ref{fig:los_PDF}, 
we compare the density contrast of pixels, each 4 pkpc in length, in the same range from 0.1-2 pMpc between our CR simulation and CROC at $z=6.5$. 
We display the density contrast Probability Density Function (PDF) for the lines of sight drawn from the CR simulation (blue curve) and that for the CROC (orange curve) in Fig.~\ref{fig:los_PDF}. 
It can be seen that our lines of sight have many more pixels with high density: the fraction of points with $\Delta_{\mathrm{g}} > 100$ is one order of magnitude larger than that in CROC. This difference is probably because the quasar host halo in the CR simulation, which reaches a mass of $ \sim 10^{13}\ M_{\odot}$ at $z=6$, is much more massive than the halos selected in \cite{Chen2021_qso_prixmity_zone}, whose halo masses are $\gtrsim 1.5\times 10^{11} M_{\odot}$.
In fact, the halo chosen in this work is more massive than almost all the halos in previous simulations of proximity zones, for example, $M_{\mathrm{h}} =2.5\times 10^{12}\ h^{-1}M_{\odot}$ in \cite{Keating2015}, 
$M_{\mathrm{h}} 
\gtrsim 10^{11.5}\ M_{\odot}$ in \cite{Davies2020_NR_RT_model}, and $M_{\mathrm{h}} < 10^{12}\ M_{\odot}$ in \cite{Satyavolu2023_qso_growth}.
We test how this difference in the surrounding density field affects the resultant $R_{\mathrm{p}}$ in Appendix~\ref{section: results: lightbulb compare} using a constant lightbulb model.
Considering that black holes were added by hand at the center of the halo in most previous studies, our lines of sight, which are self-consistently drawn from the host halo of the black hole particle, reflect the environment of the high-redshift quasar more realistically.

\subsection{Quasar light curve}
\label{section:qso_lc}

To convert the bolometric luminosity $L_{\mathrm{bol}}$ computed in Section~\ref{section:simu_phy} to the UV ionizing photon number emitted by the quasar per second ($\dot{N}$), we follow the standard procedure \cite[see, also][]{Chen2021_qso_prixmity_zone} and use a power-law spectral energy distribution (SED) from 1450 \AA\ to 912 \AA: $L_{\nu} \propto \nu^{-\alpha}$ with a spectral index $\alpha =1.5$, normalized by  $L_{\mathrm{bol}}$. This leads to $M_{\mathrm{UV}}$ upon applying the appropriate bolometric correction \citep{Fontanot2012_bol_corr}:
\begin{equation}
    M_{\mathrm{UV}} = -2.5\log_{10}\frac{L_{\mathrm{bol}}}{f_{\mathrm{B}}\mu_{\mathrm{B}}} + \Delta_{\mathrm{B,\,UV}}+ 34.1,
\end{equation}
where $f_{\mathrm{B}} = 10.2$, $\mu_{\mathrm{B}}=6.7\times10^{14}\,\mathrm{Hz}$, and $\Delta_{\mathrm{B,\,UV}} = -0.48$.
We assume the escape fraction for the quasar is $f_{\mathrm{esc}} = 100\%$, consistent with the large escape fractions inferred from observations \citep{Eiler2021_young_qso, stevans_2014_qso_fesc, Worseck2014_qso_fesc}. The total ionizing photon rate is then given by $\dot{N} = \int_{13.6\,\mathrm{eV}}^{\infty}L_{\nu}/h\nu\,d\nu$, which translates $M_{\mathrm{UV}} = -26.66$ to $\dot{N} = 1\times 10^{57}\,\text{s}^{-1}$, and $M_{\mathrm{UV}} = -27$ to $\dot{N} = 1.36 \times 10^{57}\,\text{s}^{-1}$.

We compute  $\dot{N}$ for the quasar and show the light curve (black curve) in Fig. \ref{fig:qso_LC}, compared with $\dot{N}_{\mathrm{Edd}}$ (red curve), which represents the photon number rate corresponding to $2\dot{M}_{\mathrm{Edd}}$, the upper limit of the accretion rate. The yellow curve indicates the photon number rate averaged using a $5\ \mathrm{Myr}$ top-hat kernel ($\dot{N}_{\mathrm{mean}}$).
In this work, we focus on the light curve within $6 < z \leqslant 7.5$, a period during which the average quasar luminosity ($\dot{N}_{\mathrm{mean}}$) reaches a plateau.
The quasar has a mean magnitude of $M_{\mathrm{UV,mean}}=-24.8$ in this redshift range, which is comparable to currently observed quasars. 

The quasar exhibits significant variation in the light curve as opposed to maintaining a fixed $\dot{N}$. 
We display the PDF (left panel) and the dimensionless power spectrum $kP(k)$ (right panel) of $\dot{N}/\dot{N}_{\mathrm{mean}}$ in 
Fig.~\ref{fig:qso_LC_stat}, which shows that the quasar experiences variation in $\dot{N}$ spanning over two orders of magnitude.
The characteristic fluctuation timescale, which is indicated by the peak of the power spectrum, is around $t=1$ Myr. 

\subsection{Radiative transfer code}
To interpret the effect of quasar radiation on the surrounding IGM, we carry out one-dimensional RT in the manner of \cite{Chen2021_qso_prixmity_zone}. This is implemented by postprocessing the CR simulation. 
The RT code solves the time-dependent ionization and recombination of \hi, \hei, \heii\, including the effect of quasar photoionizing radiation and the cosmic ionizing background. Temperature evolution is also calculated considering recombination cooling, collisional ionization cooling, collisional excitation cooling, Bremsstrahlung cooling, and inverse Compton cooling, as well as the expansion of the Universe. One improvement in the code compared to previous work \citep{Bolton2007_NR_RT_model, Davies2020_NR_RT_model} is the implementation of an adaptive prediction-correction scheme, which is motivated by the vastly different temporal behavior of gas at different distances from the quasar. 

The quasar light curve from the simulation is passed to the first cell, and the neutral fraction \hi, \hei, \heii\ and temperature are evolved with an adaptive scheme. At each adaptive time step, the transmitted ionizing spectrum is passed to the next cell as the incidental spectrum to evolve the next cell.
These operations are executed iteratively for consecutive cells along the line of sight. For a more complete explanation of the RT code, we direct readers to \cite{Chen2021_qso_prixmity_zone}. 

In order to compute the \Lya flux spectra, we convolve the absorption contributed by all cells with the approximate Voigt profile proposed by \cite{Tepper-Garcia2006_Voigt_profile}. This profile is calculated using the hydrogen neutral fraction $x_{\mathrm{\hi}}$ and gas temperature output from the RT simulation, in conjunction with the velocity and the density field from the CR simulation. 

\section{Results}
\label{section: results}

\begin{figure*}
	\includegraphics[width=0.9\textwidth]{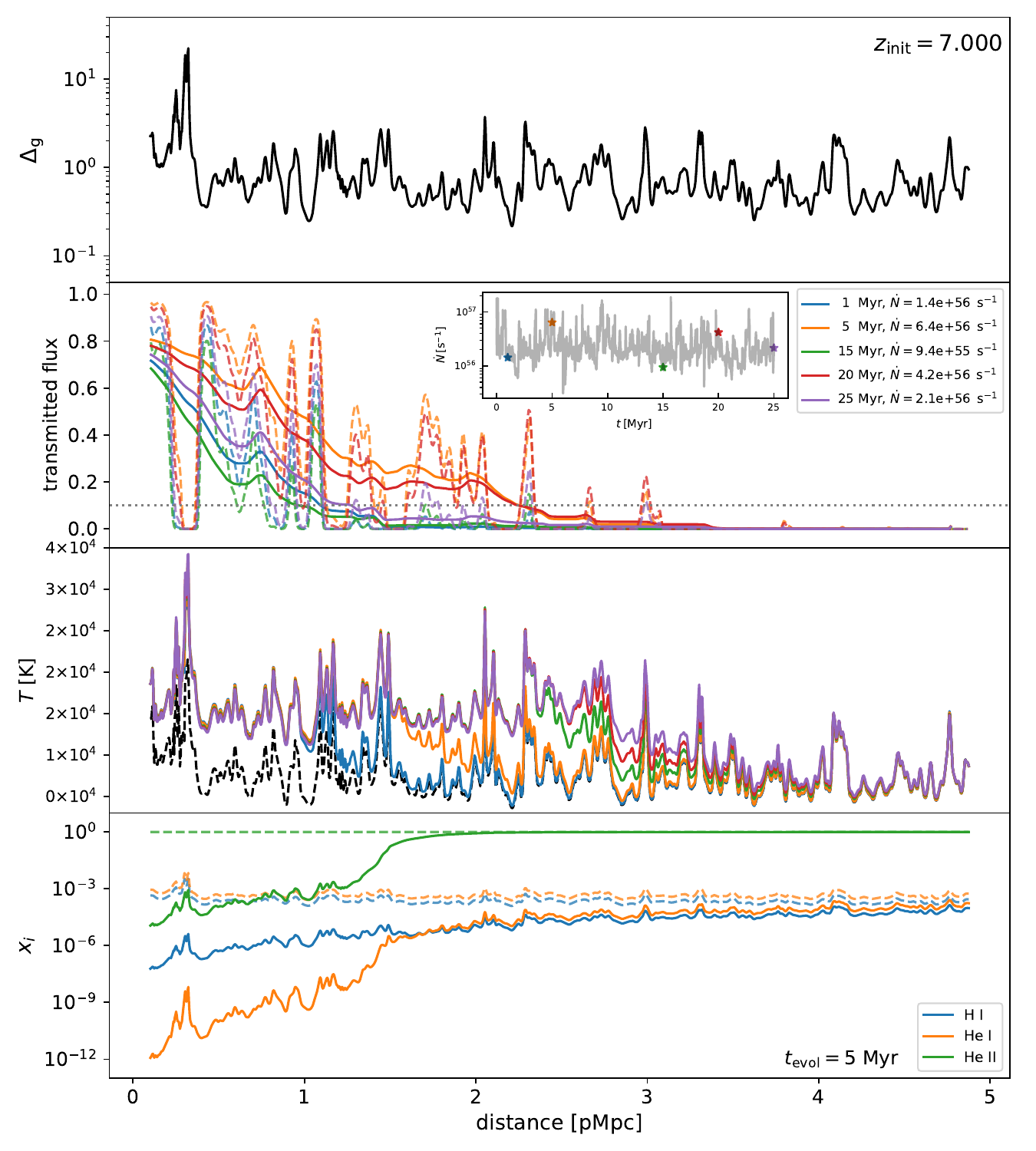}
    \caption{ The spectra computed by the RT post-processing code for one line of sight drawn from the snapshot at $z=7$. 
    Panels from top to bottom show the gas density contrast $\Delta_{\mathrm{g}}$ ($=\rho/\bar{\rho}$), the \Lya transmitted flux, the gas temperature, and the fraction of \hi/\hei/\heii. 
    The same color scheme is used in both the flux and temperature panels to illustrate the outputs from the RT simulation at $t_\mathrm{evol} = 1\ \mathrm{Myr}$ (blue), $5 \ \text{Myr}$ (orange), $15 \text{Myr}$ (green), $20\ \text{Myr}$ (red), and $25\ \text{Myr}$ (purple).
    In the second row, a subplot provides an overview of the light curve along with the $\dot{N}$ for the selected $t_{\mathrm{evol}}$ marked with colored stars, whose values are enumerated in the legend.  
    The solid curves represent the spectra smoothed by a 20 \AA\ top-hat kernel, with the dashed curves giving the original flux profiles, and the horizontal dotted line demonstrates the 10\% flux threshold. 
    In the third panel, the black dashed curve depicts the temperature at $t_\mathrm{evol} = 0\ \text{Myr}$ (i.e., no quasar radiation). 
    In the bottom row, \hi/\hei/\heii\ fraction (solid blue/orange/green curves) at $t_{\mathrm{evol}}=5\,\text{Myr}$ are compared with the background ionization fractions(dashed curves). 
    }
    \label{fig:spectra}
\end{figure*}


\begin{figure*}
	\includegraphics[width=\textwidth]{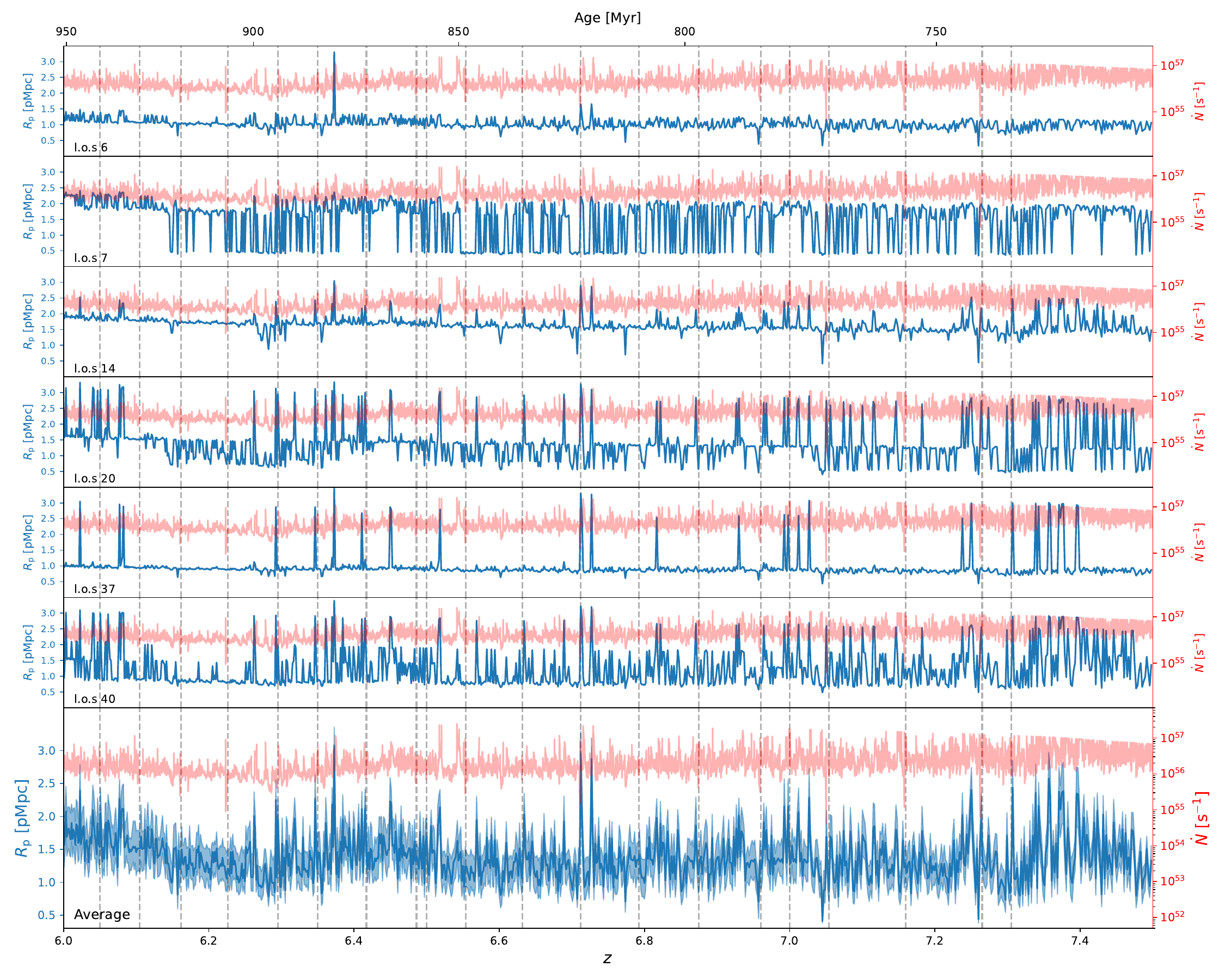}
    \caption{ The evolution of the proximity zone size $R_{\mathrm{p}}$ 
    with a time resolution of $\Delta t_{\mathrm{evol}}=0.25\ \mathrm{Myr}$.
    The top 6 panels depict the $R_{\mathrm{p}}$ (blue curve; left y-axis) for 6 lines of sight, compared with the quasar light curve $\dot{N}$ (red curves, right y-axis). 
    The vertical grey dash lines denote the redshift of available snapshots $z_{\mathrm{snap,i}}$, and the upper axis indicates the age of the Universe at a given redshift.
    The bottom panel shows the average proximity zone $\left<R_{\mathrm{p}}\right>$ (blue curve) and the 16-84th percentile scatter $\sigma_{\mathrm{R}_{\mathrm{p}}}$ (blue shaded area) 
    across 48 directions for the specific redshift.
    The standard deviation of $\left<R_{\mathrm{p}}\right>$, which represents the scatter caused by the light curve variation, is $0.33\ \mathrm{Mpc}$. And the mean of $\sigma_{\mathrm{R}_{\mathrm{p}}}$ for the entire redshift range, indicating the influence of density fluctuations, is $0.28\ \mathrm{Mpc}$.
    }
    \label{fig:long_lc_ave}
\end{figure*}

\subsection{Evolution of $R_{\mathrm{p}}$}
\label{section: results: proxmity zone evolution }

\begin{figure}
	\includegraphics[width=\columnwidth]{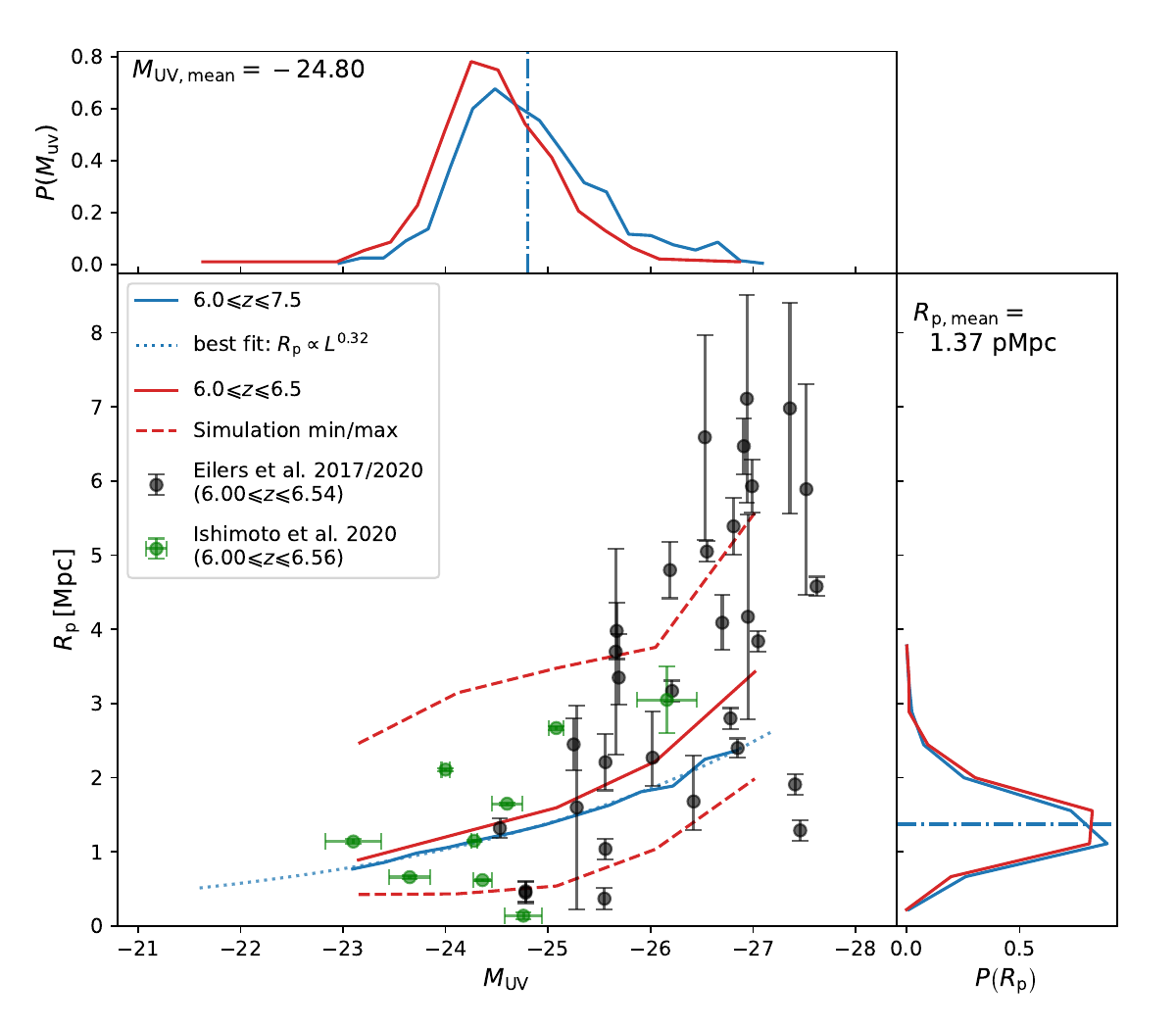}
    \caption{The dependence of proximity zone size $R_{\mathrm{p}}$ on quasar instantaneous magnitude $M_{\mathrm{UV}}$.
    Solid curves in the lower left panel show median $R_{\mathrm{p}}$ as a function of  $M_{\mathrm{UV}}$ for $6.0 \leqslant z\leqslant 7.5$ (blue), which is the entire redshift range for this study, and $6.0\leqslant z \leqslant 6.5$ (red). 
    The blue dotted curve indicates the best power-law fit $R_{\mathrm{p}}-M_{\mathrm{UV}}$ scaling relation: $R_{\mathrm{p}}\propto L^{\mathrm{0.32}}$. 
    Observational measurements from \protect\cite{Eilers2017_qso_obs}, \protect\cite{Eilers2020_qso_obs} and  \protect\cite{Ishimoto2020_qso_ob} are displayed by the black and green dots, which are made for quasars at $6.0 \leqslant z \lesssim 6.5$. The minimum and maximum simulated values for $6.0 \leqslant z \lesssim 6.5$ are depicted with the red dashed curves for comparison.
    The upper and the right panels present the marginal distribution of $M_{\mathrm{UV}}$ and $R_{\mathrm{p}}$, respectively. The blue dash-dot lines represent the mean values of the $M_{\mathrm{UV}}$ and the $R_{\mathrm{p}}$ for $6.0 \leqslant z\leqslant 7.5$: $M_{\mathrm{UV,mean}}=-24.8$,\ $R_{\mathrm{p,mean}}=1.37$ pMpc.}
    \label{fig:R_dist}
\end{figure}

We present the post-processed spectra obtained from the RT code for a single line of sight at $z=7$
in Fig. \ref{fig:spectra}.  The gas density contrast $\Delta_{\mathrm{g}}$ is shown in the first row
and the temperature prior to quasar activation (i.e., $t_{\mathrm{evol}}=0\,\text{Myr}$) is shown in the third row with the black dashed line. 
The transmitted flux and IGM temperature at  $t_{\mathrm{evol}}=1\,\text{Myr}$ (blue), 
$5\,\text{Myr}$ (orange), 
$15\,\text{Myr}$ (green), 
$20\,\text{Myr}$ (red), 
and $25\,\text{Myr}$ (purple) are depicted in the second and third rows, respectively, with the corresponding $\dot{N}$ values enumerated in the legend of the second panel. A subplot in the second row provides an overview of the light curve variation along with the selected $\dot{N}$ marked by stars. 
The traditional definition of the proximity zone edge is the point where the \Lya flux first drops below 10\% after being smoothed by a 20\AA\ top-hat window. 
We show the smoothed flux (solid curves) as well as the flux threshold 10\% (horizontal grey dashed line) in the second panel, and the $R_{\mathrm{p}}$ is indicated by the intersection of the horizontal line and the solid curves. 
It is noteworthy that the higher the stars marked in the light curve, i.e., larger instantaneous $\dot{N}$, the higher the corresponding flux is. 
This implies that the levels of the \Lya flux, and consequently the proximity zone size $R_{\mathrm{p}}$, are primarily determined by $\dot{N}$ while showing no correlation with $t_{\mathrm{evol}}$.

The profiles of \hi/\hei/\heii\ fractions (solid blue/orange/green curves) at $t_{\mathrm{evol}}=5\,\text{Myr}$ are displayed in the bottom row, compared with the background profile (dashed lines). 
In the original CR simulation, helium is predominantly found in the \heii\ state. As the quasar's radiation ionizes the \heii, a `\heii\ proximity zone' emerges, and the energy injected from the \heii\ reionization heats the surrounding IGM, which is known as the `thermal proximity effect' \citep{Bolton2010_thermal_effect, Bolton2012_thermal_effect, Meiksin2010_thermal_effect}. 
This phenomenon also enhances the size of the \Lya proximity zone since the \hi\ fraction is partially temperature-dependent \citep{Davies2020_NR_RT_model}. 
As observed in the third panel of Fig.~\ref{fig:spectra}, the heated region, unlike $R_{\mathrm{p}}$, continues to expand monotonically with the increasing $t_{\mathrm{evol}}$, suggesting its potential application in the estimation of quasar lifetimes (see Section~\ref{section: qso_age}).

In order to fully exploit the 21 snapshots within the redshift range: $6 < z_{\mathrm{snap, i}} \leqslant 7.5\  (1 \leqslant i  \leqslant 21;\ z_{\mathrm{snap, i+1}}<z_{\mathrm{snap, i}})$, we conduct an RT calculation on the $i\mathrm{th}$ snapshot for a period of 
\begin{equation}
    t_{\mathrm{evol, i}} = t_{\mathrm{age}}(z_{\mathrm{snap,i+1}}) - t_{\mathrm{age}}(z_{\mathrm{snap,i}}),
\end{equation}
where $t_{\mathrm{age}}(z)$ is the age of the universe at redshift $z$, $z_{\mathrm{snap,i+1}}$ is the redshift of the subsequent snapshot and $z_{\mathrm{snap,22}}=6$. 
When we concatenate the next snapshot at $z_{\mathrm{snap,i}}\ (i>1)$, we draw the density and velocity along the lines of sight from the new snapshot, which has not been post-processed, while using the temperature and the ionization fraction of \hi/\hei/\heii\ output by the RT code from the previous snapshot.
We then concatenate these evolution segments, each reflecting the $R_{\mathrm{p}}$ during $t_{\mathrm{age}}(z_{\mathrm{snap,i}}) \sim t_{\mathrm{age}}(z_{\mathrm{snap, i+1}})$ interval.
Such a stitching operation is implemented across all the lines of sight, each maintaining a fixed direction throughout the entire redshift range. 
We estimate the uncertainty in the resultant $R_{\mathrm{p}}$ caused by the breaking of the continuous evolution of the IGM  caused by this procedure in Appendix~\ref{section: error_ana}. We find it tiny compared to the $R_{\mathrm{p}}$ scatter caused by the underlying density fluctuations.

We capture the variations in proximity zone sizes across 48 directions with a variable light curve for $\sim 240\ \mathrm{Myr}$ after the quasar's activation.
Fig. \ref{fig:long_lc_ave} presents the evolution of $R_{\mathrm{p}}$ with a time resolution of $\Delta t_{\mathrm{evol}}=0.25\ \mathrm{Myr}$ for 6 sample sightlines (blue curves in the upper six panels) as well as an average value for 48 directions (blue curve in the bottom row).
As a comparison, we also plot the quasar light curves $\dot{N}$ using the red curves.
The blue shaded area in the bottom panel depicts the 16-84th percentile scatter among different directions. 
The vertical grey dash lines indicate the redshifts of available snapshots $z_{\mathrm{snap,i}}$.
Since the reionization is virtually completed at $z=7.5$, the scatter between different lines of sight is entirely attributable to the underlying density field at a specific point in time \citep{Lidz2006_los_scatter}. 
On the other hand, the fluctuations in the quasar light curve contribute significantly to the variability in $R_{\mathrm{p}}$ for an individual direction within short time frames. The extent of this variability in $R_{\mathrm{p}}$ depends on the specific density field.
This is evidenced by a comparison of the proximity zone evolution between sightlines 6 and 7 (the second and third rows in Fig. \ref{fig:long_lc_ave}): the same light curve fluctuations result in changes of $\Delta R_{\mathrm{p}}\sim 1.7\ \mathrm{Mpc}$ for sightline 7 within $0.25\ \mathrm{Myr}$, while for sightline 6, the changes are  nearly all less than $\Delta R_{\mathrm{p}}\sim 0.5\ \mathrm{Mpc}$.

We can quantify the influence of the underlying density fluctuations by computing the standard deviation of proximity zone sizes ($\sigma_{R_{\mathrm{p}}}$) across sightlines for the same redshift. We find the mean of this standard deviation ($\left<\sigma_{R_{\mathrm{p}}}\right>$) is $0.28\ \mathrm{Mpc}$ across the entire redshift range. 
On the other hand, to show the influence of light curve variability, we compute the mean of \Rp $(\left<\mathrm{R}_{\mathrm{p}}\right>)$ across all sightlines at a specific redshift, and then calculate the standard deviation of these mean \Rp values for different redshifts ($\sigma_{\left<\mathrm{R}_{\mathrm{p}}\right>}$), which is $0.33\ \mathrm{Mpc}$.
 This illustrates that compared to the density fluctuations, the variations in the light curve have a similar influence, or
 slightly larger, on the scatter of $R_{\mathrm{p}}$ values.

\begin{figure*}
\includegraphics[width=\textwidth]
{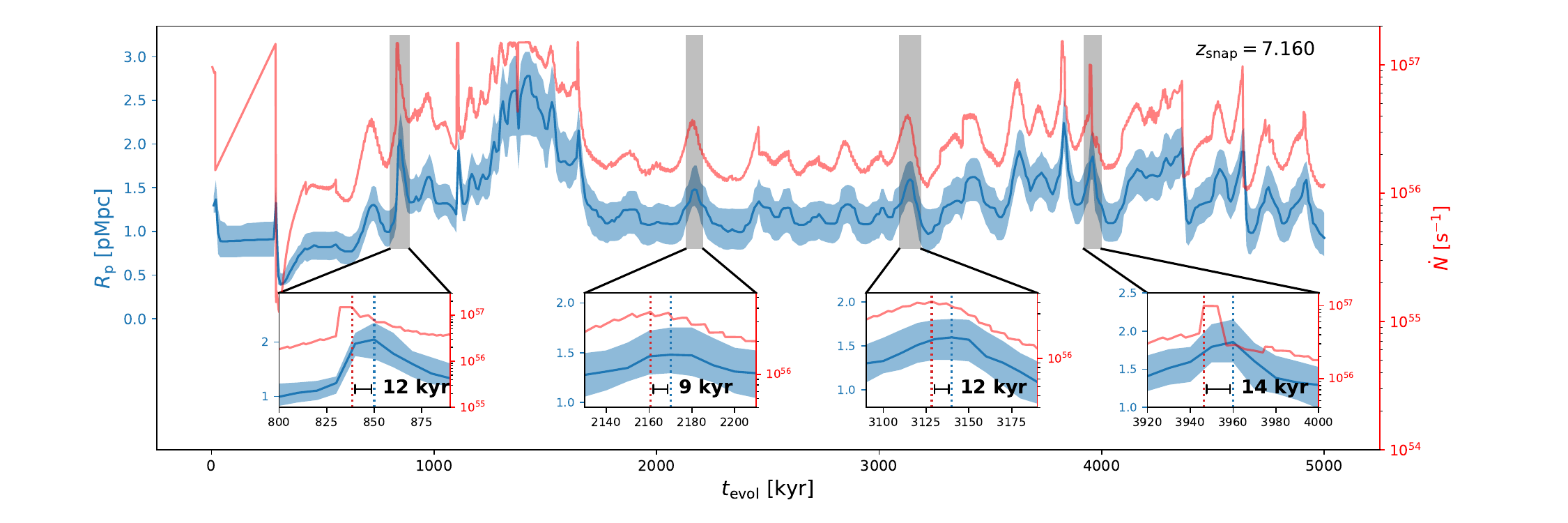}
    \caption{The averaged evolution of the proximity zone size $R_{\mathrm{p}}$ (blue curve; left y-axis) compared with the quasar light curve $\dot{N}$ (red curve; right y-axis) for $5\ \mathrm{Myr}$ following $z_{\mathrm{snap}} = 7.160$. The blue shaded region represents the 16-84th percentile scatter. The subplots zoom in to $80\ \mathrm{kyr}$ time spans around four peaks in the light curve (red dotted lines in subplots) and their corresponding $R_{\mathrm{p}}$ peaks (blue dotted lines in subplots), which are denoted by the black shaded rectangles. In the subplots, we label the time lag between the light curve peaks and the $R_{\mathrm{p}}$ peaks, which are $\sim 10^{4}\ \mathrm{yr}$.
}
    \label{fig:Robs_evol_zoomin}
\end{figure*}

\subsection{$R_{\mathrm{p}}\, - M_{\mathrm{UV}}$ scaling relation}
\label{section: proximity zone distribution}

In Fig. \ref{fig:R_dist}, the lower left panel displays the proximity zone size $R_{\mathrm{p}}$ as a function of the quasar's instantaneous magnitude $M_{\mathrm{UV}}$.
The median $R_{\mathrm{p}}$ across the entire redshift range ($6.0\leqslant z \leqslant 7.5$) for 48 lines of sight is represented by the blue solid curve.
The best power-law fit $R_{\mathrm{p}}-M_{\mathrm{UV}}$ scaling relation is found to be $\log R_{\mathrm{p}}\propto -0.13\,M_{\mathrm{uv}}$, i.e.,  $R_{\mathrm{p}}\propto L^{0.32}$ (blue dotted curve). The upper and the right panels show the marginal distribution of $M_{\mathrm{UV}}$ and $R_{\mathrm{p}}$.
The blue dash-dot lines represent the mean values of  $M_{\mathrm{UV}}$ and  $R_{\mathrm{p}}$: $M_{\mathrm{UV,mean}}=-24.8$,\ $R_{\mathrm{p,mean}}=1.37$ pMpc.
The $M_{\mathrm{UV}}$ histogram shows that the quasar is relatively faint ($M_{\mathrm{UV}}\gtrsim -25$) over most of its lifetime, and only becomes luminous ($M_{\mathrm{UV}}\lesssim -26$) occasionally. 

For comparison with observational results, in Fig.\ref{fig:R_dist} we plot the observed proximity zone size for quasars in the range  $6\leqslant z\lesssim 6.5$ measured by \cite{Eilers2017_qso_obs, Eilers2020_qso_obs} (black dots) and  \cite{Ishimoto2020_qso_ob} (green dots).
We also show the minimum/maximum and median values yielded by our simulation within the same redshift range using red dashed lines and a solid line, respectively. 
Our simulation reproduces the $R_{\mathrm{p}}$ range for most of the observed quasars with $M_{\mathrm{UV}}>-26$. The inability to yield $R_{\mathrm{p}} < 0.5\ \mathrm{Mpc}$ could be due to the failure to resolve LLSs or DLAs. On the bright end, our simulation predicts smaller $R_{\mathrm{p}}$ than some of the observational measurements.
This probably stems from the quasar's low average luminosity,
which we discuss in more detail in Section~\ref{section: lc_effect}.

Our derived $R_{\mathrm{p}}-M_{\mathrm{UV}}$ scaling relation exhibits a flatter trend compared to the $R_{\mathrm{p}}\propto L^{0.5}$ predicted by \cite{Bolton2007_NR_RT_model} for an idealized ionized IGM based on a semi-analytical model (see equation~\ref{equ:ana_Rp_BH2007}). 
One possible reason for the disparity is that the $M_{\mathrm{UV}}$ for our simulated quasar is non-uniformly distributed across a broad redshift range, as shown in the upper panel of Fig.~\ref{fig:R_dist}. The $R_{\mathrm{p}}-M_{\mathrm{UV}}$ scaling relation is redshift-dependent, which can be seen in Fig.~\ref{fig:lightbulb_test}, where the lower redshift environment typically produces more extensive proximity zones.
According to our simulation, the optimal fit is 
$\log R_{\mathrm{p}} \approx \log(3.20\ \mathrm{Mpc})  + 0.36\left[-0.4(M_{\mathrm{UV}}+27)\right]$
 at $6.0 \leqslant z \leqslant 6.5$ and $\log R_{\mathrm{p}} \approx \log(2.41\ \mathrm{Mpc})  + 0.41\left[-0.4(M_{\mathrm{UV}}+27)\right]$ at $7.3 \leqslant z \leqslant 7.5$. 
The slope remains shallower than that in \cite{Bolton2007_NR_RT_model}, even when the data is constrained within a narrower redshift band.
Such a weaker dependence of $R_{\mathrm{p}}$ on the instantaneous magnitude probably originates from the variation in the light curve, which breaks the correspondence between the $R_{\mathrm{p}}$ and the contemporaneous $M_{\mathrm{UV}}$ as we discuss in Section~\ref{section: qso_age}.

\section{Discussion}
\label{section:discuss}

\subsection{Response of \Rp to variable light curve}
\label{section: qso_age}

Several recent studies have focused on constraining quasar lifetimes using proximity zone sizes under the assumption of a lightbulb model \citep{Morey2021_qso_tq_nz, Khrykin2021_qso_tq}. In this section, we briefly discuss the influence of a variable light curve on quasar lifetime estimation.

We start by exploring the response behavior of $R_{\mathrm{p}}$ to the quasar light curve. We plot a $5\ \mathrm{Myr}$ duration in the evolution of $R_{\mathrm{p}}$ starting from $z_{\mathrm{snap}}=7.160$ in Fig. \ref{fig:Robs_evol_zoomin}, with the blue curve representing the average value for 48 lines of sight and the shaded area showing the 16-84th percentile scatter. 
The subplots zoom in to $80\ \mathrm{kyr}$ time spans around four peaks in the light curve (red dotted lines in subplots) and their corresponding $R_{\mathrm{p}}$ peaks (blue dotted lines in subplots). 
We label the time lags between the light curve peaks and the $R_{\mathrm{p}}$ peaks in each subplot, which are $12,\ 9,\ 12,\ 14\ $ kyr, respectively. They are comparable to the hydrogen equilibrium time at the edge of the proximity zone: $t_{\mathrm{eq}}^{\mathrm{\hi}} =1/\Gamma^{\mathrm{\hi}} \sim 10^{4}$ yr, where $\Gamma^{\mathrm{\hi}}$ is the photoionization rate of hydrogen \citep{Bolton2007, Eilers2018, Davies2020_NR_RT_model}.
This illustrates that $R_{\mathrm{p}}$ traces the fluctuations in the light curve closely but with a short delay of $\sim 10^{4}$ yr, which breaks the correspondence between  $R_{\mathrm{p}}$ and the contemporaneous $M_{\mathrm{UV}}$.

Previous studies have extensively discussed the implications of quasar proximity zone size for their ‘lifetime,’ using a lightbulb model. This model describes a quasar turning on suddenly, with its luminosity remaining constant thereafter. Our simulated light curve features some episodes where the luminosity increases nearly by a factor of $\times 4$ within $10^3$ years (e.g., at 830 kyr in the first zoom-in panel and at 3945 kyr in the last zoom-in panel, as shown in Fig. \ref{fig:Robs_evol_zoomin}). Therefore, these sudden jumps in luminosity can be viewed as the beginning of an `episode', a term previous studies have used to describe the quasar’s episodic lifetime \citep{Eilers2017_qso_obs,Eiler2021_young_qso}. However, we note that the light curve varies rapidly and seldom behaves like a lightbulb for more than a few times $10^3$ years.
By the time $10^4$ yr have passed, denoted as the typical $R_{\mathrm{p}}$ delay time, the quasar’s luminosity has already changed significantly. Moreover, there are many periods during which the quasar luminosity evolves slowly (e.g., as seen in the second and third zoom-in panels of Fig. \ref{fig:Robs_evol_zoomin}), making the ‘episodic lifetime’ ill-defined.

In the latter half of reionization, the integrated lifetime of the quasar can hardly be measured solely based on the size of the proximity zone. The $R_{\mathrm{p}}$ value only informs us about the quasar’s luminosity within a span of $10^4$ years (as seen in Fig. \ref{fig:Robs_evol_zoomin}), while the integrated lifetime of our quasar has been hundreds of million years. To measure this total duration for which the quasar has been shining, one can use observable associated with the thermal states of the IGM around the quasar, like the He II proximity zone, as it has a longer response time $t_{\mathrm{eq}} \sim 10^{6}\ \mathrm{yr}$ \citep{Worseck2021_qso_dating_heii, Khrykin2017_qso_tq_heii, Khrykin2021_qso_tq,Soltinsky2023_qso_lifetime,chen2023}.

\begin{figure*}
	\includegraphics[width=\textwidth]{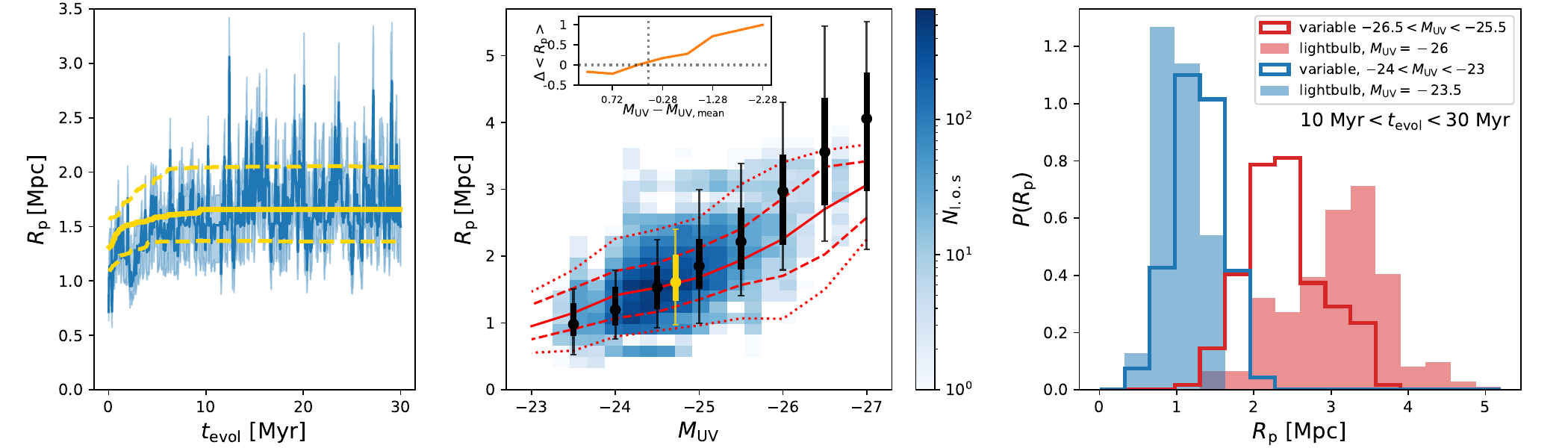}
\caption{
Left panel: the evolution of the mean proximity zone sizes $R_{\mathrm{p}}$ with the variable light curve (blue curves) or a lightbulb model fixed at $M_{\mathrm{UV,mean}}=-24.72$ (yellow curves).
The quasar is assumed to turn on at $z=6.162$ and last for $30\ \mathrm{Myr}$. 
The blue shaded region and the yellow dashed lines indicate the 68\% scatter for the two $R_{\mathrm{p}}$ groups.
Middle panel: two-dimensional distribution of $R_{\mathrm{p}}$ and instantaneous $M_{\mathrm{UV}}$ for $10<t_{\mathrm{evol}}<30$ Myr. The blue pixels represent the $R_{\mathrm{p}}$ for the variable light curve, whose median/68\%/95\% confidence regions are shown by the red solid/dashed/dotted curves. The dots depict the $R_{\mathrm{p}}$ generated by the lightbulb models fixing the magnitude at $M_{\mathrm{UV}}=-23.5,\ -24,\ -24.5,\ -25,\ -25.5,\  -26,\ -26.5,\ -27$ (black dots) and $M_{\mathrm{UV,mean}}=-24.74$ (yellow dot).
The thick/thin errorbars correspond to 68\%/95\% confidence intervals.
The subplot displays the difference between the median $R_{\mathrm{p}}$ produced by the lightbulb models and the variable light curve. The horizontal dotted line represents where the two median values are the same, and the vertical dotted line shows $M_{\mathrm{UV, mean}}$.
Right panel: one-dimensional $R_{\mathrm{p}}$ distribution generated by the part of the variable light curve with $-26.5 < M_{\mathrm{UV}} <  -25.5$ (red curve), $-24 < M_{\mathrm{UV}} <  -23$ (blue curve), and by a lightbulb model with $M_{\mathrm{UV}}=-26$ (red shaded area), $M_{\mathrm{UV}}=-23.5$ (blue shaded area) for $10\ \mathrm{Myr}<t_{\mathrm{evol}}<30\ \mathrm{Myr}$.
}\label{fig:LC_compare}
\end{figure*}

\subsection{Influence of light curve variation}
\label{section: lc_effect}

In this section, we discuss how the rapid fluctuation in the light curve in our simulation affects the resultant $R_{\mathrm{p}}$, 
and compare it with the proximity zone sizes yielded by a light curve that remains constant over an extended period of time.

We excerpt a portion of the variable light curve from the simulation at $z=6.162$ with a  time span of $30\ \mathrm{Myr}$, short enough that the density evolution is negligible. This excerpted light curve has a mean magnitude of $M_{\mathrm{UV,mean}}=-24.72$ and a mean ionizing photon rate of $\dot{N}_{\mathrm{mean}} = 1.67 \times 10^{56} \ \mathrm{s}^{-1}$.  In contrast to this variable light curve, we construct another light curve with a constant flux of the same $\dot{N}_{\mathrm{mean}}$ (`lightbulb' model). 
We evolve the same set of $48$ sightlines with these two light curves for $30$ Myr.
In the left panel of Fig.~\ref{fig:LC_compare}, we display the mean $R_{\mathrm{p}}$ at different $t_{\mathrm{evol}}$ of the variable light curve (blue curve) and the lightbulb model (solid yellow curve).
The blue shaded region and the yellow dashed lines indicate the 68\% scatter for the two $R_{\mathrm{p}}$ groups.
For the lightbulb model, $R_{\mathrm{p}}$ remains nearly unchanged after rapid growth during the first $\sim 1\ \mathrm{Myr}$, with only a slight decline owing to the Universe cooling, which aligns with the results of previous studies \citep{Davies2020_NR_RT_model, Eiler2021_young_qso}.
In the middle panel, we show the $R_{\mathrm{p}}$ distributions as a function of instantaneous magnitude $M_{\mathrm{UV}}$ for $10<t_{\mathrm{evol}}<30$ Myr, during which the lightbulb model reaches a stable stage and produces a similar \Rp.
The blue pixels represent the $R_{\mathrm{p}}$ for the variable 
light curve, whose median, 68\%, and 95\% scatter regions are shown by the red solid, dashed, and dotted curves, respectively.
The $R_{\mathrm{p}}$ values generated by the lightbulb models are  shown as dots, whose thick and thin errorbars correspond to the 68\% and 95\% scatter, respectively.
With a similar mean proximity zone size $\left<R_{\mathrm{p}}\right>\sim 1.6 \ \mathrm{pMpc}$ across the entire magnitude range, 
the variable light curve yields a scatter ($\sigma_{\mathrm{R_{\mathrm{p}}}}=0.46\ \mathrm{pMpc}$) 28\% larger than that of the $\dot{N}_{\mathrm{mean}}$ lightbulb ($\sigma_{\mathrm{R_{\mathrm{p}}}}=0.36\ \mathrm{pMpc}$).
The $\sigma_{\mathrm{R_{\mathrm{p}}}}$ values for the variable light curve encompass contributions from both light curve variation and underlying density field fluctuation. 
On the other hand, the $\sigma_{\mathrm{R_{\mathrm{p}}}}$ for the lightbulb model is almost totally attributed to the density differences. 
In addition to the lightbulb model fixed at $M_{\mathrm{UV,mean}}$ (yellow dot), we also simulate the lightbulb with different magnitudes (black dots): $M_{\mathrm{UV}}=-23.5,\ -24,\ -24.5,\ -25,\ -25.5,\  -26,\ -26.5,\ -27$ in the middle panel.
It can be seen from the error bars that the influence of the density fluctuation for a lightbulb is strongly correlated with $M_{\mathrm{UV}}$, and the high luminosity magnifies the variance between directions, i.e., brighter $M_{\mathrm{UV}}$ leads to an increase in $\sigma_{R_\mathrm{p}}$.

An important feature illustrated by the middle panel of Fig.~\ref{fig:LC_compare} is that the median $R_{\mathrm{p}}$ from the variable light curve at a specific magnitude coincides with the lightbulb model only around $M_{\mathrm{UV, mean}}=-24.72$, while it tends to  yield smaller $R_{\mathrm{p}}$ compared to the lightbulb when $M_{\mathrm{UV}} < M_{\mathrm{UV, mean}}$, and conversely, larger $R_{\mathrm{p}}$ when $M_{\mathrm{UV}} > M_{\mathrm{UV, mean}}$. 
This is more clearly demonstrated in the subplot, where we 
show the difference between the median $R_{\mathrm{p}}$ produced by the lightbulb models and the variable light curve. The horizontal dotted line represents where the two median values are the same, and the vertical dotted line shows $M_{\mathrm{UV, mean}}$.
More specifically, in the right panel of Fig.~\ref{fig:LC_compare} we compare the one-dimensional $R_{\mathrm{p}}$ distributions within a narrow $M_{\mathrm{UV}}$ bin for $10\ \mathrm{Myr}<t_{\mathrm{evol}}<30\ \mathrm{Myr}$.
On the bright end ($M_{\mathrm{UV}}\sim -26$), the lightbulb model predicts a median $R_{\mathrm{p}}$ 30\% larger than that produced by the variable light curve ($2.94$ pMpc versus $2.27$ pMpc).
While on the dim end ($M_{\mathrm{UV}}\sim -23.5$), the lightbulb model yields a median $R_{\mathrm{p}}$ 13\% smaller than that of the variable light curve ($1.03$ pMpc versus $1.19$ pMpc).
However, these two models give similar scatter in $R_{\mathrm{p}}$  for this quasar. 
The standard deviations of $R_{\mathrm{p}}$ for both the variable light curve and the lightbulb model are  $\sim 0.6$ pMpc around $M_{\mathrm{UV}} = -26$, and $\sim 0.3$ pMpc around $M_{\mathrm{UV}}= -23.5$. 

By building a toy model of the fluctuating light curves, \cite{Davies2020_NR_RT_model} noticed that the $R_{\mathrm{p}}$ simulated based on variable light curves skewed towards smaller values as opposed to a lightbulb fixed at a relatively high luminosity.
A similar bias on the bright end emerges in our simulation, while  
our computation herein further demonstrates that the discrepancy between the lightbulb model and the variable light curve is contingent upon the specific magnitude bin. 
Such a discrepancy occurs because $R_{\mathrm{p}}$ is governed by the entire light curve within the most recent $\sim 10^{4}\ \mathrm{yr}$, rather than the contemporaneous instantaneous luminosity. 
As depicted by the PDF in Fig.~\ref{fig:qso_LC_stat}, the distribution of $\dot{N}$ is essentially Gaussian centering around $\dot{N}_{\mathrm{mean}}$, which implies that the $\dot{N}$ value $10^{4}\ \mathrm{yr}$ preceding a bright or a dim point in the light curve is probably close to $\dot{N}_{\mathrm{mean}}$, producing a $R_{\mathrm{p}}$ close to that given by a lightbulb model fixed at $\dot{N}_{\mathrm{UV,mean}}$. 
Furthermore, the higher luminosities correspond to more significant variations since the variation amplitude generally equals $M_{\mathrm{UV,mean}} - M_{\mathrm{UV}}$. Large variations result in more remarkable discrepancies (see Appendix~\ref{app: var_mag}), which explains the more substantial shift at smaller magnitudes observed in the middle panel of Fig.~\ref{fig:LC_compare}.

Therefore, for an individual quasar whose light curve persistently fluctuates around a certain value, its proximity zone size displays a shallow evolution with instantaneous magnitude, and diverts from the lightbulb model in a $M_{\mathrm{UV}}$-dependent way. 
Such divergence accounts for the difference between our predicted $R_{\mathrm{p}}$ and the observational measurements at $M_{\mathrm{UV}} <-26$ shown in Fig.~\ref{fig:R_dist}: our simulated light curve generally has a lower luminosity, which makes the $R_{\mathrm{p}}$ in this magnitude range smaller; while the observed quasars with $M_{\mathrm{UV}} <-26$ probably have larger overall luminosity, and so generate large proximity zones.

\begin{figure}
	\includegraphics[width=\columnwidth]{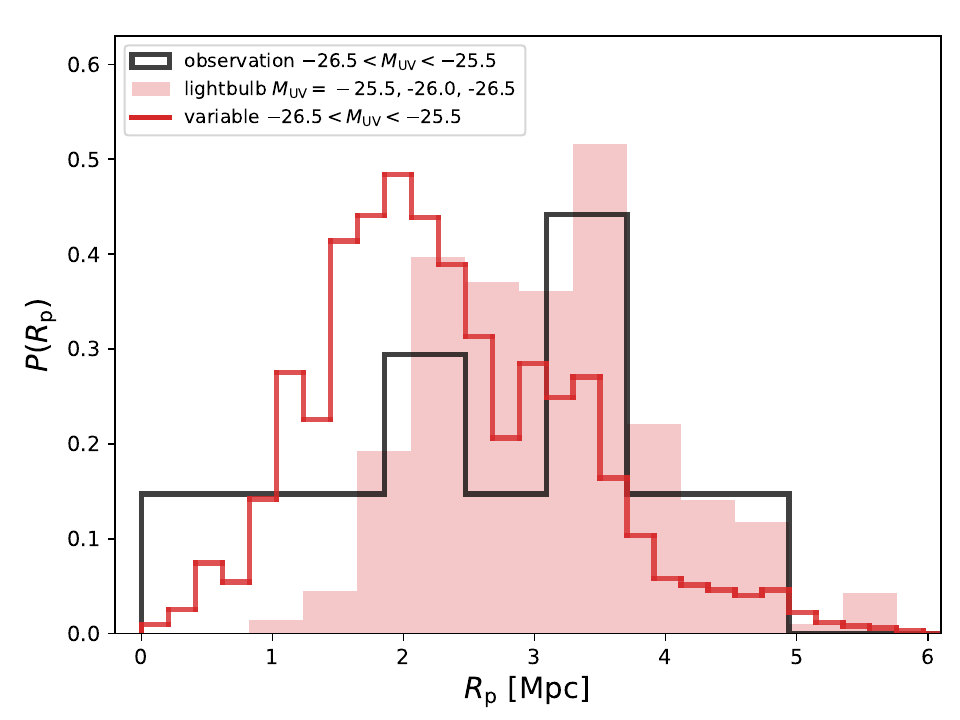}
    \caption{The $R_{\mathrm{p}}$ distribution with $-26.5<M_{\mathrm{UV}}<-25.5$ produced by the variable light curves (red solid curve), and the lightbulb models with $10<t_{\mathrm{evol}}<30$ Myr (red shaded area), compared with the observed $R_{\mathrm{p}}$ for the quasars within the same magnitude bin measured in \protect\cite{Eilers2017_qso_obs, Eilers2020_qso_obs, Ishimoto2020_qso_ob}.
    } 
    \label{fig:hist_obs_26}
\end{figure}

\subsection{Implications of the observed \protect$R_{\mathrm{p}}$ distribution}
\label{section: Rp_dis_obs}

Our simulation shows that one single quasar can vary significantly over its lifetime, with a scatter in luminosity spanning approximately two orders of magnitude. If we define it according to its mean luminosity \MeanMuv, our quasar is a relatively faint one (\Muv=-24.8) during the period $z=7.5\sim6$. However, it still has a $16\%$  chance to be caught in a relatively bright phase with \Muv<-25.5. In such a bright phase, the distribution of \Rp is shifted towards the shorter end compared to the case of constant luminosity (the right panel of Fig. \ref{fig:LC_compare}). This has profound implications for interpreting the \Rp distribution at given observed magnitude \Muv ($P(\rp|\muv)$).

The distribution of $P(\rp|\muv)$ can be formulated as the following conditional distribution:
$$ P(\rp|\muv) = \frac{P(\meanMuv)P(\muv|\meanMuv)P(\rp|\muv,\meanMuv)}{P(\meanMuv)P(\muv|\meanMuv)},$$
where $P(\meanMuv)$ is the probability function of quasars with a certain mean magnitude \MeanMuv, $P(\muv|\meanMuv)$ is the probability that a quasar of \MeanMuv is observed with magnitude \Muv, and $P(\rp|\muv,\meanMuv)$ is the probability that the quasar of \MeanMuv at the observed magnitude \Muv displays a proximity zone size of \Rp.

To calculate such a distribution, certain assumptions need to be made.
The first term $P(\meanMuv)$ is similar to the quasar luminosity function, but it is for the mean magnitude \MeanMuv instead of the observed luminosity. Measuring such a \MeanMuv directly is challenging. For simplicity, here we assume that $P(\meanMuv)$ is equal to the observed quasar luminosity function (QLF) measured by \cite{Matsuoka2018_qso_LF} for quasars at $z=6$:
\begin{equation}
    P(M_{\mathrm{UV}}^{\star}) \propto \left[10^{-0.156\left(M_{\mathrm{UV}}^{\star} + 25.30\right)} + 10^{-0.716\left(M_{\mathrm{UV}}^{\star} + 25.30\right)}\right]^{-1}.
    \label{equ:qlf}
\end{equation}

To estimate $P(M_{\mathrm{UV}}|M_{\mathrm{UV}}^{\star})$, we need to know the PDF of the light curve. Motivated by the luminosity PDF of our simulated quasar (upper panel of Fig.~\ref{fig:R_dist}), we assume that the light curve has a Gaussian distribution centered at $M_{\mathrm{UV}}^{\star}$:
\begin{equation}
    P(M_{\mathrm{UV}}|M_{\mathrm{UV}}^{\star}) = \frac{1}{\sigma\sqrt{2\pi}}\,\exp\left[\frac{1}{2}\left(\frac{M_{\mathrm{UV}}-M_{\mathrm{UV}}^{\star}}{\sigma}\right)^{2}\right].
\end{equation}
with a fixed scatter $\sigma=0.7$ for all the light curves.

Finally, to model 
$P(R_{\mathrm{p}} |M_{\mathrm{UV}},M_{\mathrm{UV}}^{\star})$,
we make the following assumptions: (1) $P(R_{\mathrm{p}} |M_{\mathrm{UV}},M_{\mathrm{UV}}^{\star})$ has the same shape as the $R_{\mathrm{p}}$ distribution generated by the constant light curve with magnitude fixed at $M_{\mathrm{UV}}$, which we label as $P_{\mathrm{lb},M_{\mathrm{UV}}}(R_{\mathrm{p}})$, but is shifted towards a different mean $R_{\mathrm{p}}$ with the amount $B$. 
(2) the value of $B$ is only determined by $M_{\mathrm{UV}} - M_{\mathrm{UV}}^{\star}$ and is independent of the specific values of $M_{\mathrm{UV}}^{\star}$ and $M_{\mathrm{UV}}$. 
Hence, the probability of $R_{\mathrm{p}}$ for a given variable light curve is
\begin{equation}    
P(R_{\mathrm{p}} |M_{\mathrm{UV}},M_{\mathrm{UV}}^{\star}) = P_{\mathrm{lb},M_{\mathrm{UV}}}(R_{\mathrm{p}} + B).
\end{equation}
The first assumption is motivated by the right panel of Fig.~\ref{fig:LC_compare}, which shows that for an individual quasar, the $R_{\mathrm{p}}$ distributions generated by the variable light curve and the lightbulb have similar shapes: they have roughly the same scatter but  different mean $R_{\mathrm{p}}$. To formulate $B(M_{\mathrm{UV}} - M_{\mathrm{UV}}^{\star})$, we use our results in Section.~\ref{section: lc_effect} as a guideline, i.e.,
we linearly interpolate the difference between the median $R_{\mathrm{p}}$ of the lightbulb models and the variable light curve as a function of the magnitude difference $M_{\mathrm{UV}} - M_{\mathrm{UV}}^{\star}$, which is depicted by the subplot in the middle panel of Fig.~\ref{fig:LC_compare}.

With the formulas and assumptions stated above, we use Markov Chain Monte Carlo (MCMC) to generate $N=1000$ quasar samples to create the final distribution $P(\rp|\muv)$.
We present the $R_{\mathrm{p}}$ distribution with $-26.5<M_{\mathrm{UV}}<-25.5$ calculated by this model in Fig.~\ref{fig:hist_obs_26} (red solid curve).
As a comparison, we plot the combined $R_{\mathrm{p}}$ distribution produced by the lightbulb models for $10<t_{\mathrm{evol}}<30\ \mathrm{Myr}$
fixed at $M_{\mathrm{UV}}=-25.5,\ -26,\ -26.5$ (red shaded area). We consider three lightbulb models rather than only the one with $M_{\mathrm{UV}}=-26$ to show the whole $R_{\mathrm{p}}$ range reached by the lightbulb in this $M_{\mathrm{UV}}$ bin, which spans the minimum $R_{\mathrm{p}}$, reached when $M_{\mathrm{UV}}=-25.5$, to the maximum $R_{\mathrm{p}}$, for $M_{\mathrm{UV}}=-26.5$.
These three lightbulb models are sampled based on the QLF (equation~\ref{equ:qlf}).
We also show the $R_{\mathrm{p}}$ values for the observed quasars with $-26.5<M_{\mathrm{UV}}<-25.5$ measured by \cite{Eilers2017_qso_obs, Eilers2020_qso_obs, Ishimoto2020_qso_ob} in Fig.~\ref{fig:hist_obs_26} (black curve).
It is evident that although a variable light curve produces a similar maximum value ($R_{\mathrm{p, max}}\sim 5.5$ Mpc) as the lightbulb model, it can also result in much smaller proximity zones ($R_{\mathrm{p, min}}\sim 0.2\ \mathrm{Mpc}$). Therefore, it readily explains the observations with small proximity zones. These extremely small $R_{\mathrm{p}}$ values are generated by the quasar with low averaged luminosity (i.e., large $M_{\mathrm{UV}}^{\star}$). 
The considerable difference between the observed instantaneous $M_{\mathrm{UV}}$, which is $\sim -26$ in this case, and $M_{\mathrm{UV}}^{\star}$ moves the $R_{\mathrm{p}}$ distribution away from the distribution without light curve variation, as seen in the right panel of Fig. \ref{fig:LC_compare}. Additionally, due to the relative abundance of faint $M_{\mathrm{UV}}^{\star}$ quasars over the bright ones, the overall $R_{\mathrm{p}}$ distribution skews towards smaller values.

Since the calculation for the variable light curve and the lightbulb models are based on the same groups of lines of sight, and because the lightbulb model accounts for all the scatter introduced by underlying density fluctuations, the wider scatter shown by the red curve in Fig.~\ref{fig:hist_obs_26} can only be attributed to light curve variability. 
This underscores the necessity of considering light curve variability when investigating quasar proximity zones. 

Note that \cite{Davies2020_NR_RT_model} made a similar comparison in their Fig. 16 between the observed \Rp distribution and predicted \Rp from different quasar light curve models. They concluded that the lightbulb model with long episodic quasar lifetime ($\geq 1$ Myr) leads to a $R_{\mathrm{p}}$ distribution consistent with the observed  \Rp distribution, while their toy quasar light curve with variation does not. 
On the contrary, our simulated quasar light curve results in a \Rp distribution that skews only slightly to the smaller end compared to the lightbulb model (red line versus transparent red shaded histograms in Fig. \ref{fig:hist_obs_26}). We conduct a Kolmogorov–Smirnov (K-S) test and find the K-S statistic to be $D=0.23$ and $p$-value$=0.54$ when comparing the observed \Rp distribution and that from the lightbulb model. On the other hand, we find $D=0.34$ and $p$-value$=0.12$ between the observed \Rp distribution and the one from our variable light curve.
Therefore, both the lightbulb model and our simulated variable light curve are compatible with the observed \Rp distribution.
We reach this different conclusion from \cite{Davies2020_NR_RT_model} because (1) their toy quasar light curve is constructed to have large variability at very small time scales ($10^2$ yr and $10^4$ yr), while our simulated quasar light curve has low power at such small timescales (see Fig. \ref{fig:qso_LC_stat}); (2) we consider the combined $R_{\mathrm{p}}$ distribution generated by a group of variable quasars with different mean magnitude, rather than using an individual quasar; (3) we use an observational sample from \cite{Eilers2017_qso_obs, Eilers2020_qso_obs, Ishimoto2020_qso_ob} while \cite{Davies2020_NR_RT_model}  only have the data from \cite{Eilers2017_qso_obs}.

As discussed in this subsection, the distribution of \Rp for a given observed \Muv bin is influenced by both the underlying quasar mean-luminosity function and quasar variability, which includes both the luminosity PDF and the light curve power spectrum. Measuring \Rp
can therefore provide constraints on these critical properties of the first quasars. However, it is important to note that the current observed \Rp
distribution may be incomplete, and the selection function is complex. As a result, comparisons between models and observed \Rp should be interpreted cautiously. Obtaining a large complete sample of \Rp for reionization-era quasars could significantly enhance our understanding of quasar variability. From a modeling perspective, future work will explore how different light curve power spectra affect the
\Rp distribution.

\section{Conclusions}
\label{section:conclusion}
In this work, we study the proximity zone around a high-redshift quasar through RT post-processing the lines of sight from a cosmological simulation with constrained initial conditions. This constrained realization creates a quasar host halo of $M_{h}=10^{13}\ M_{\odot}$ at $z=6$, more massive than most halos studied in previous simulations. The simulation also includes galaxy and black hole formation models, resulting in a variable quasar light curve, which is more realistic than the widely used lightbulb model. The simulated light curve exhibits extreme variability, with the changes in luminosity spanning up to two orders of magnitude around the average value.

By concatenating $R_{\mathrm{p}}$ evolution segments from 21 snapshots covering $6.0< z \leqslant 7.5$, we capture the variations in proximity zone sizes with a variable light curve for $\sim 240\ \mathrm{Myr}$ after the quasar's activation in a highly ionized IGM.  The resultant $R_{\mathrm{p}}$ ranges between $0.5-5\ \mathrm{pMpc}$. 
We demonstrate that variations in the light curve contribute an additional scatter, which is separate from the scatter induced by density variations. The standard deviation in $R_{\mathrm{p}}$  values caused by each of these effects are approximately \( \sigma(R_{\mathrm{p}}) \sim 0.3\ \mathrm{pMpc} \).

Our simulation suggests that in a pre-ionized IGM, the evolution of $R_{\mathrm{p}}$ traces the variations in the light curve closely with a short time delay of $\sim 10^{4}\ \mathrm{yr}$.
This time lag breaks the correspondence between the $R_{\mathrm{p}}$ and the contemporaneous $M_{\mathrm{UV}}$. 
The $R_{\mathrm{p}}$ is heavily influenced by the magnitude about $10^{4}$ yr previously, whose difference from the observed $M_{\mathrm{UV}}$ is uncertain and could be significant.This indicates that $R_{\mathrm{p}}$ can only be used to infer the quasar episodic lifetime at best, and does not inform us of the integrated quasar lifetime. 

By analyzing the $R_{\mathrm{p}}$ distribution for specific $M_{\mathrm{UV}}$ values,
we show that for an individual quasar with a fluctuating light curve,  its proximity zone size increases weakly with brighter instantaneous magnitude, and diverts from the lightbulb model in a $M_{\mathrm{UV}}$-dependent way.
Compared to the variable light curve, the lightbulb model underestimates $R_{\mathrm{p}}$ by 13\% at the dim end ($M_{\mathrm{UV}}\sim -23.5$), and overestimates the $R_{\mathrm{p}}$ by 30\% at the bright end ($M_{\mathrm{UV}}\sim -26$). 



We computed the distribution of $R_{\mathrm{p}}$ based on a set of quasars sampled from a QLF and found that light curve variability leads to a broad distribution of $R_{\mathrm{p}}$ at given observed magnitude.
Notably, variable light curves contribute to a group of instantaneously bright quasars with extremely small proximity zones. 
These small $R_{\mathrm{p}}$ can hardly be explained if the quasar light curve stays constant. This shows that it is  necessary to consider the details of light curve variability when investigating quasar proximity zones.

\section*{Acknowledgements}
The authors thank Hy Trac and Nianyi Chen for helpful discussions. 
HC thanks the support by the Natural Sciences and Engineering Research Council of Canada (NSERC), funding reference \#DIS-2022-568580.
SB acknowledges the funding support by NASA-80NSSC22K1897.
TDM and RAAC acknowledge funding from the NSF AI Institute: Physics of the Future, NSF PHY-2020295, NASA ATP NNX17AK56G, and NASA ATP 80NSSC18K101. TDM acknowledges additional support from NASA ATP 19-ATP19-0084, and NASA ATP 80NSSC20K0519.

\section*{Data Availability}

The data underlying this article will be shared on reasonable request to the corresponding author.



\bibliographystyle{mnras}
\bibliography{reference} 

\begin{thebibliography}{}
\makeatletter
\relax
\def\mn@urlcharsother{\let\do\@makeother \do\$\do\&\do\#\do\^\do\_\do\%\do\~}
\def\mn@doi{\begingroup\mn@urlcharsother \@ifnextchar [ {\mn@doi@} {\mn@doi@[]}}
\def\mn@doi@[#1]#2{\def\@tempa{#1}\ifx\@tempa\@empty \href {http://dx.doi.org/#2} {doi:#2}\else \href {http://dx.doi.org/#2} {#1}\fi \endgroup}
\def\mn@eprint#1#2{\mn@eprint@#1:#2::\@nil}
\def\mn@eprint@arXiv#1{\href {http://arxiv.org/abs/#1} {{\tt arXiv:#1}}}
\def\mn@eprint@dblp#1{\href {http://dblp.uni-trier.de/rec/bibtex/#1.xml} {dblp:#1}}
\def\mn@eprint@#1:#2:#3:#4\@nil{\def\@tempa {#1}\def\@tempb {#2}\def\@tempc {#3}\ifx \@tempc \@empty \let \@tempc \@tempb \let \@tempb \@tempa \fi \ifx \@tempb \@empty \def\@tempb {arXiv}\fi \@ifundefined {mn@eprint@\@tempb}{\@tempb:\@tempc}{\expandafter \expandafter \csname mn@eprint@\@tempb\endcsname \expandafter{\@tempc}}}

\bibitem[\protect\citeauthoryear{{Ba{\~n}ados} et~al.,}{{Ba{\~n}ados} et~al.}{2018}]{Banados2018_qso_75}
{Ba{\~n}ados} E.,  et~al., 2018, \mn@doi [\nat] {10.1038/nature25180}, \href {https://ui.adsabs.harvard.edu/abs/2018Natur.553..473B} {553, 473}

\bibitem[\protect\citeauthoryear{{Bajtlik}, {Duncan}  \& {Ostriker}}{{Bajtlik} et~al.}{1988}]{Bajtlik1988}
{Bajtlik} S.,  {Duncan} R.~C.,   {Ostriker} J.~P.,  1988, \mn@doi [\apj] {10.1086/166217}, \href {https://ui.adsabs.harvard.edu/abs/1988ApJ...327..570B} {327, 570}

\bibitem[\protect\citeauthoryear{{Battaglia}, {Trac}, {Cen}  \& {Loeb}}{{Battaglia} et~al.}{2013}]{Battaglia2013_patchy_reion}
{Battaglia} N.,  {Trac} H.,  {Cen} R.,   {Loeb} A.,  2013, \mn@doi [\apj] {10.1088/0004-637X/776/2/81}, \href {https://ui.adsabs.harvard.edu/abs/2013ApJ...776...81B} {776, 81}

\bibitem[\protect\citeauthoryear{{Bolton} \& {Haehnelt}}{{Bolton} \& {Haehnelt}}{2007a}]{Bolton2007_NR_RT_model}
{Bolton} J.~S.,  {Haehnelt} M.~G.,  2007a, \mn@doi [\mnras] {10.1111/j.1365-2966.2006.11176.x}, \href {https://ui.adsabs.harvard.edu/abs/2007MNRAS.374..493B} {374, 493}

\bibitem[\protect\citeauthoryear{{Bolton} \& {Haehnelt}}{{Bolton} \& {Haehnelt}}{2007b}]{Bolton2007}
{Bolton} J.~S.,  {Haehnelt} M.~G.,  2007b, \mn@doi [\mnras] {10.1111/j.1745-3933.2007.00361.x}, \href {https://ui.adsabs.harvard.edu/abs/2007MNRAS.381L..35B} {381, L35}

\bibitem[\protect\citeauthoryear{{Bolton}, {Becker}, {Wyithe}, {Haehnelt}  \& {Sargent}}{{Bolton} et~al.}{2010}]{Bolton2010_thermal_effect}
{Bolton} J.~S.,  {Becker} G.~D.,  {Wyithe} J. S.~B.,  {Haehnelt} M.~G.,   {Sargent} W. L.~W.,  2010, \mn@doi [\mnras] {10.1111/j.1365-2966.2010.16701.x}, \href {https://ui.adsabs.harvard.edu/abs/2010MNRAS.406..612B} {406, 612}

\bibitem[\protect\citeauthoryear{{Bolton}, {Haehnelt}, {Warren}, {Hewett}, {Mortlock}, {Venemans}, {McMahon}  \& {Simpson}}{{Bolton} et~al.}{2011}]{Bolton2011_qso_7_IGM}
{Bolton} J.~S.,  {Haehnelt} M.~G.,  {Warren} S.~J.,  {Hewett} P.~C.,  {Mortlock} D.~J.,  {Venemans} B.~P.,  {McMahon} R.~G.,   {Simpson} C.,  2011, \mn@doi [\mnras] {10.1111/j.1745-3933.2011.01100.x}, \href {https://ui.adsabs.harvard.edu/abs/2011MNRAS.416L..70B} {416, L70}

\bibitem[\protect\citeauthoryear{{Bolton}, {Becker}, {Raskutti}, {Wyithe}, {Haehnelt}  \& {Sargent}}{{Bolton} et~al.}{2012}]{Bolton2012_thermal_effect}
{Bolton} J.~S.,  {Becker} G.~D.,  {Raskutti} S.,  {Wyithe} J. S.~B.,  {Haehnelt} M.~G.,   {Sargent} W. L.~W.,  2012, \mn@doi [\mnras] {10.1111/j.1365-2966.2011.19929.x}, \href {https://ui.adsabs.harvard.edu/abs/2012MNRAS.419.2880B} {419, 2880}

\bibitem[\protect\citeauthoryear{Bondi \& Hoyle}{Bondi \& Hoyle}{1944}]{Bondi_Hoyle_1944_accretion_rate}
Bondi H.,  Hoyle F.,  1944, \mn@doi [MNRAS] {10.1093/mnras/104.5.273}, 104, 273

\bibitem[\protect\citeauthoryear{{Bosman} \& {Becker}}{{Bosman} \& {Becker}}{2015a}]{Bosman2015_qso_7_IGM}
{Bosman} S. E.~I.,  {Becker} G.~D.,  2015a, \mn@doi [\mnras] {10.1093/mnras/stv1336}, \href {https://ui.adsabs.harvard.edu/abs/2015MNRAS.452.1105B} {452, 1105}

\bibitem[\protect\citeauthoryear{{Bosman} \& {Becker}}{{Bosman} \& {Becker}}{2015b}]{Bosman2015_11200641}
{Bosman} S. E.~I.,  {Becker} G.~D.,  2015b, \mn@doi [\mnras] {10.1093/mnras/stv1336}, \href {https://ui.adsabs.harvard.edu/abs/2015MNRAS.452.1105B} {452, 1105}

\bibitem[\protect\citeauthoryear{{Carilli} et~al.,}{{Carilli} et~al.}{2010}]{Carilli2010_qso_ob}
{Carilli} C.~L.,  et~al., 2010, \mn@doi [\apj] {10.1088/0004-637X/714/1/834}, \href {https://ui.adsabs.harvard.edu/abs/2010ApJ...714..834C} {714, 834}

\bibitem[\protect\citeauthoryear{{Cen} \& {Haiman}}{{Cen} \& {Haiman}}{2000}]{Cen2000}
{Cen} R.,  {Haiman} Z.,  2000, \mn@doi [\apjl] {10.1086/312937}, \href {https://ui.adsabs.harvard.edu/abs/2000ApJ...542L..75C} {542, L75}

\bibitem[\protect\citeauthoryear{{Chen} \& {Gnedin}}{{Chen} \& {Gnedin}}{2021}]{Chen2021_qso_prixmity_zone}
{Chen} H.,  {Gnedin} N.~Y.,  2021, \mn@doi [\apj] {10.3847/1538-4357/abe7e7}, \href {https://ui.adsabs.harvard.edu/abs/2021ApJ...911...60C} {911, 60}

\bibitem[\protect\citeauthoryear{{Chen}, {Croft}  \& {Gnedin}}{{Chen} et~al.}{2023}]{chen2023}
{Chen} H.,  {Croft} R. A.~C.,   {Gnedin} N.~Y.,  2023, \mn@doi [\mnras] {10.1093/mnras/stad049}, \href {https://ui.adsabs.harvard.edu/abs/2023MNRAS.519.5931C} {519, 5931}

\bibitem[\protect\citeauthoryear{{Davies} et~al.,}{{Davies} et~al.}{2018}]{Davies2018}
{Davies} F.~B.,  et~al., 2018, \mn@doi [\apj] {10.3847/1538-4357/aad6dc}, \href {https://ui.adsabs.harvard.edu/abs/2018ApJ...864..142D} {864, 142}

\bibitem[\protect\citeauthoryear{{Davies}, {Hennawi}  \& {Eilers}}{{Davies} et~al.}{2020}]{Davies2020_NR_RT_model}
{Davies} F.~B.,  {Hennawi} J.~F.,   {Eilers} A.-C.,  2020, \mn@doi [\mnras] {10.1093/mnras/stz3303}, \href {https://ui.adsabs.harvard.edu/abs/2020MNRAS.493.1330D} {493, 1330}

\bibitem[\protect\citeauthoryear{{DeGraf}, {Di Matteo}, {Treu}, {Feng}, {Woo}  \& {Park}}{{DeGraf} et~al.}{2015}]{DeGraf2015_simu_obs_test}
{DeGraf} C.,  {Di Matteo} T.,  {Treu} T.,  {Feng} Y.,  {Woo} J.~H.,   {Park} D.,  2015, \mn@doi [\mnras] {10.1093/mnras/stv2002}, \href {https://ui.adsabs.harvard.edu/abs/2015MNRAS.454..913D} {454, 913}

\bibitem[\protect\citeauthoryear{{Di Matteo}, {Springel}  \& {Hernquist}}{{Di Matteo} et~al.}{2005}]{DiMatteo2005_BH_model}
{Di Matteo} T.,  {Springel} V.,   {Hernquist} L.,  2005, \mn@doi [\nat] {10.1038/nature03335}, \href {https://ui.adsabs.harvard.edu/abs/2005Natur.433..604D} {433, 604}

\bibitem[\protect\citeauthoryear{{Di Matteo}, {Khandai}, {DeGraf}, {Feng}, {Croft}, {Lopez}  \& {Springel}}{{Di Matteo} et~al.}{2012}]{dimatteo2012}
{Di Matteo} T.,  {Khandai} N.,  {DeGraf} C.,  {Feng} Y.,  {Croft} R.~A.~C.,  {Lopez} J.,   {Springel} V.,  2012, \mn@doi [\apjl] {10.1088/2041-8205/745/2/L29}, \href {https://ui.adsabs.harvard.edu/abs/2012ApJ...745L..29D} {745, L29}

\bibitem[\protect\citeauthoryear{{Ding}, {Treu}, {Silverman}, {Bhowmick}, {Menci}  \& {Di Matteo}}{{Ding} et~al.}{2020}]{Ding2020_simu_obs_test}
{Ding} X.,  {Treu} T.,  {Silverman} J.~D.,  {Bhowmick} A.~K.,  {Menci} N.,   {Di Matteo} T.,  2020, \mn@doi [\apj] {10.3847/1538-4357/ab91be}, \href {https://ui.adsabs.harvard.edu/abs/2020ApJ...896..159D} {896, 159}

\bibitem[\protect\citeauthoryear{Ding et~al.,}{Ding et~al.}{2022}]{Ding_2022_simu_obs_test}
Ding X.,  et~al., 2022, \mn@doi [The Astrophysical Journal] {10.3847/1538-4357/ac714c}, 933, 132

\bibitem[\protect\citeauthoryear{{Eilers}, {Davies}, {Hennawi}, {Prochaska}, {Luki{\'c}}  \& {Mazzucchelli}}{{Eilers} et~al.}{2017}]{Eilers2017_qso_obs}
{Eilers} A.-C.,  {Davies} F.~B.,  {Hennawi} J.~F.,  {Prochaska} J.~X.,  {Luki{\'c}} Z.,   {Mazzucchelli} C.,  2017, \mn@doi [\apj] {10.3847/1538-4357/aa6c60}, \href {https://ui.adsabs.harvard.edu/abs/2017ApJ...840...24E} {840, 24}

\bibitem[\protect\citeauthoryear{{Eilers}, {Hennawi}  \& {Davies}}{{Eilers} et~al.}{2018}]{Eilers2018}
{Eilers} A.-C.,  {Hennawi} J.~F.,   {Davies} F.~B.,  2018, \mn@doi [\apj] {10.3847/1538-4357/aae081}, \href {https://ui.adsabs.harvard.edu/abs/2018ApJ...867...30E} {867, 30}

\bibitem[\protect\citeauthoryear{{Eilers} et~al.,}{{Eilers} et~al.}{2020}]{Eilers2020_qso_obs}
{Eilers} A.-C.,  et~al., 2020, \mn@doi [\apj] {10.3847/1538-4357/aba52e}, \href {https://ui.adsabs.harvard.edu/abs/2020ApJ...900...37E} {900, 37}

\bibitem[\protect\citeauthoryear{{Eilers}, {Hennawi}, {Davies}  \& {Simcoe}}{{Eilers} et~al.}{2021}]{Eiler2021_young_qso}
{Eilers} A.-C.,  {Hennawi} J.~F.,  {Davies} F.~B.,   {Simcoe} R.~A.,  2021, \mn@doi [\apj] {10.3847/1538-4357/ac0a76}, \href {https://ui.adsabs.harvard.edu/abs/2021ApJ...917...38E} {917, 38}

\bibitem[\protect\citeauthoryear{{Fan} et~al.,}{{Fan} et~al.}{2006}]{Fan2006_qso_obs}
{Fan} X.,  et~al., 2006, \mn@doi [\aj] {10.1086/504836}, \href {https://ui.adsabs.harvard.edu/abs/2006AJ....132..117F} {132, 117}

\bibitem[\protect\citeauthoryear{{Feng}, {Di-Matteo}, {Croft}, {Bird}, {Battaglia}  \& {Wilkins}}{{Feng} et~al.}{2016}]{Feng2016_Bluetides}
{Feng} Y.,  {Di-Matteo} T.,  {Croft} R.~A.,  {Bird} S.,  {Battaglia} N.,   {Wilkins} S.,  2016, \mn@doi [\mnras] {10.1093/mnras/stv2484}, \href {https://ui.adsabs.harvard.edu/abs/2016MNRAS.455.2778F} {455, 2778}

\bibitem[\protect\citeauthoryear{{Feng}, {Bird}, {Anderson}, {Font-Ribera}  \& {Pedersen}}{{Feng} et~al.}{2018}]{Feng2018_MPGadget}
{Feng} Y.,  {Bird} S.,  {Anderson} L.,  {Font-Ribera} A.,   {Pedersen} C.,  2018, {Mp-Gadget/Mp-Gadget: A Tag For Getting A Doi}, Zenodo, \mn@doi{10.5281/zenodo.1451799}

\bibitem[\protect\citeauthoryear{{Ferrara}, {Salvadori}, {Yue}  \& {Schleicher}}{{Ferrara} et~al.}{2014}]{Ferrara2014_dc_bh_seed}
{Ferrara} A.,  {Salvadori} S.,  {Yue} B.,   {Schleicher} D.,  2014, \mn@doi [\mnras] {10.1093/mnras/stu1280}, \href {https://ui.adsabs.harvard.edu/abs/2014MNRAS.443.2410F} {443, 2410}

\bibitem[\protect\citeauthoryear{{Fontanot}, {Cristiani}  \& {Vanzella}}{{Fontanot} et~al.}{2012}]{Fontanot2012_bol_corr}
{Fontanot} F.,  {Cristiani} S.,   {Vanzella} E.,  2012, \mn@doi [\mnras] {10.1111/j.1365-2966.2012.21594.x}, \href {https://ui.adsabs.harvard.edu/abs/2012MNRAS.425.1413F} {425, 1413}

\bibitem[\protect\citeauthoryear{{Gabor} \& {Bournaud}}{{Gabor} \& {Bournaud}}{2013}]{Gabor2013_qso_lc}
{Gabor} J.~M.,  {Bournaud} F.,  2013, \mn@doi [\mnras] {10.1093/mnras/stt1046}, \href {https://ui.adsabs.harvard.edu/abs/2013MNRAS.434..606G} {434, 606}

\bibitem[\protect\citeauthoryear{{Greig}, {Mesinger}, {Haiman}  \& {Simcoe}}{{Greig} et~al.}{2017}]{Greig2017_11200641}
{Greig} B.,  {Mesinger} A.,  {Haiman} Z.,   {Simcoe} R.~A.,  2017, \mn@doi [\mnras] {10.1093/mnras/stw3351}, \href {https://ui.adsabs.harvard.edu/abs/2017MNRAS.466.4239G} {466, 4239}

\bibitem[\protect\citeauthoryear{{Greig}, {Mesinger}, {Davies}, {Wang}, {Yang}  \& {Hennawi}}{{Greig} et~al.}{2022}]{Greig2022_dampingwing}
{Greig} B.,  {Mesinger} A.,  {Davies} F.~B.,  {Wang} F.,  {Yang} J.,   {Hennawi} J.~F.,  2022, \mn@doi [\mnras] {10.1093/mnras/stac825}, \href {https://ui.adsabs.harvard.edu/abs/2022MNRAS.512.5390G} {512, 5390}

\bibitem[\protect\citeauthoryear{{Haiman} \& {Cen}}{{Haiman} \& {Cen}}{2001}]{Haiman2001}
{Haiman} Z.,  {Cen} R.,  2001, in {Umemura} M.,  {Susa} H.,  eds,  Astronomical Society of the Pacific Conference Series Vol. 222, The Physics of Galaxy Formation. p.~101

\bibitem[\protect\citeauthoryear{{Hinshaw} et~al.,}{{Hinshaw} et~al.}{2013}]{Hinshaw_2013_WMAP}
{Hinshaw} G.,  et~al., 2013, \mn@doi [\apjs] {10.1088/0067-0049/208/2/19}, \href {https://ui.adsabs.harvard.edu/abs/2013ApJS..208...19H} {208, 19}

\bibitem[\protect\citeauthoryear{{Hoffman} \& {Ribak}}{{Hoffman} \& {Ribak}}{1991}]{Hoffman1991_CR}
{Hoffman} Y.,  {Ribak} E.,  1991, \mn@doi [\apjl] {10.1086/186160}, \href {https://ui.adsabs.harvard.edu/abs/1991ApJ...380L...5H} {380, L5}

\bibitem[\protect\citeauthoryear{{Hopkins}}{{Hopkins}}{2013}]{Hopkins2013_pSPH}
{Hopkins} P.~F.,  2013, \mn@doi [\mnras] {10.1093/mnras/sts210}, \href {https://ui.adsabs.harvard.edu/abs/2013MNRAS.428.2840H} {428, 2840}

\bibitem[\protect\citeauthoryear{{Inayoshi}, {Visbal}  \& {Haiman}}{{Inayoshi} et~al.}{2020}]{Inayoshi2020}
{Inayoshi} K.,  {Visbal} E.,   {Haiman} Z.,  2020, \mn@doi [\araa] {10.1146/annurev-astro-120419-014455}, \href {https://ui.adsabs.harvard.edu/abs/2020ARA&A..58...27I} {58, 27}

\bibitem[\protect\citeauthoryear{{Ishimoto} et~al.,}{{Ishimoto} et~al.}{2020}]{Ishimoto2020_qso_ob}
{Ishimoto} R.,  et~al., 2020, \mn@doi [\apj] {10.3847/1538-4357/abb80b}, \href {https://ui.adsabs.harvard.edu/abs/2020ApJ...903...60I} {903, 60}

\bibitem[\protect\citeauthoryear{{Katz}, {Weinberg}  \& {Hernquist}}{{Katz} et~al.}{1996}]{Katz1996_gas_cooling_radia}
{Katz} N.,  {Weinberg} D.~H.,   {Hernquist} L.,  1996, \mn@doi [\apjs] {10.1086/192305}, \href {https://ui.adsabs.harvard.edu/abs/1996ApJS..105...19K} {105, 19}

\bibitem[\protect\citeauthoryear{{Keating}, {Haehnelt}, {Cantalupo}  \& {Puchwein}}{{Keating} et~al.}{2015}]{Keating2015}
{Keating} L.~C.,  {Haehnelt} M.~G.,  {Cantalupo} S.,   {Puchwein} E.,  2015, \mn@doi [\mnras] {10.1093/mnras/stv2020}, \href {https://ui.adsabs.harvard.edu/abs/2015MNRAS.454..681K} {454, 681}

\bibitem[\protect\citeauthoryear{{Khandai}, {Di Matteo}, {Croft}, {Wilkins}, {Feng}, {Tucker}, {DeGraf}  \& {Liu}}{{Khandai} et~al.}{2015}]{Khandai2015_massiveBH}
{Khandai} N.,  {Di Matteo} T.,  {Croft} R.,  {Wilkins} S.,  {Feng} Y.,  {Tucker} E.,  {DeGraf} C.,   {Liu} M.-S.,  2015, \mn@doi [\mnras] {10.1093/mnras/stv627}, \href {https://ui.adsabs.harvard.edu/abs/2015MNRAS.450.1349K} {450, 1349}

\bibitem[\protect\citeauthoryear{{Khrykin}, {Hennawi}  \& {McQuinn}}{{Khrykin} et~al.}{2017}]{Khrykin2017_qso_tq_heii}
{Khrykin} I.~S.,  {Hennawi} J.~F.,   {McQuinn} M.,  2017, \mn@doi [\apj] {10.3847/1538-4357/aa6621}, \href {https://ui.adsabs.harvard.edu/abs/2017ApJ...838...96K} {838, 96}

\bibitem[\protect\citeauthoryear{{Khrykin}, {Hennawi}, {Worseck}  \& {Davies}}{{Khrykin} et~al.}{2021}]{Khrykin2021_qso_tq}
{Khrykin} I.~S.,  {Hennawi} J.~F.,  {Worseck} G.,   {Davies} F.~B.,  2021, \mn@doi [\mnras] {10.1093/mnras/stab1288}, \href {https://ui.adsabs.harvard.edu/abs/2021MNRAS.505..649K} {505, 649}

\bibitem[\protect\citeauthoryear{{King} \& {Nixon}}{{King} \& {Nixon}}{2015}]{King2015_qso_lc}
{King} A.,  {Nixon} C.,  2015, \mn@doi [\mnras] {10.1093/mnrasl/slv098}, \href {https://ui.adsabs.harvard.edu/abs/2015MNRAS.453L..46K} {453, L46}

\bibitem[\protect\citeauthoryear{{Kravtsov}}{{Kravtsov}}{1999}]{kravtsov1999}
{Kravtsov} A.~V.,  1999, PhD thesis, NEW MEXICO STATE UNIVERSITY

\bibitem[\protect\citeauthoryear{{Kravtsov}, {Klypin}  \& {Hoffman}}{{Kravtsov} et~al.}{2002}]{kravtsov2002}
{Kravtsov} A.~V.,  {Klypin} A.,   {Hoffman} Y.,  2002, \mn@doi [\apj] {10.1086/340046}, \href {https://ui.adsabs.harvard.edu/abs/2002ApJ...571..563K} {571, 563}

\bibitem[\protect\citeauthoryear{{Krumholz} \& {Gnedin}}{{Krumholz} \& {Gnedin}}{2011}]{Krumholz2011_sfr}
{Krumholz} M.~R.,  {Gnedin} N.~Y.,  2011, \mn@doi [\apj] {10.1088/0004-637X/729/1/36}, \href {https://ui.adsabs.harvard.edu/abs/2011ApJ...729...36K} {729, 36}

\bibitem[\protect\citeauthoryear{{Latif}, {Schleicher}, {Schmidt}  \& {Niemeyer}}{{Latif} et~al.}{2013}]{Latif2013_dc_bh_seed}
{Latif} M.~A.,  {Schleicher} D.~R.~G.,  {Schmidt} W.,   {Niemeyer} J.~C.,  2013, \mn@doi [\mnras] {10.1093/mnras/stt1786}, \href {https://ui.adsabs.harvard.edu/abs/2013MNRAS.436.2989L} {436, 2989}

\bibitem[\protect\citeauthoryear{{Li}}{{Li}}{2012}]{Li2012_SMBH_acc}
{Li} L.-X.,  2012, \mn@doi [\mnras] {10.1111/j.1365-2966.2012.21336.x}, \href {https://ui.adsabs.harvard.edu/abs/2012MNRAS.424.1461L} {424, 1461}

\bibitem[\protect\citeauthoryear{{Lidz}, {Oh}  \& {Furlanetto}}{{Lidz} et~al.}{2006}]{Lidz2006_los_scatter}
{Lidz} A.,  {Oh} S.~P.,   {Furlanetto} S.~R.,  2006, \mn@doi [\apjl] {10.1086/502678}, \href {https://ui.adsabs.harvard.edu/abs/2006ApJ...639L..47L} {639, L47}

\bibitem[\protect\citeauthoryear{{Lidz}, {McQuinn}, {Zaldarriaga}, {Hernquist}  \& {Dutta}}{{Lidz} et~al.}{2007}]{Lidz2007}
{Lidz} A.,  {McQuinn} M.,  {Zaldarriaga} M.,  {Hernquist} L.,   {Dutta} S.,  2007, \mn@doi [\apj] {10.1086/521974}, \href {https://ui.adsabs.harvard.edu/abs/2007ApJ...670...39L} {670, 39}

\bibitem[\protect\citeauthoryear{Liu \& Liu}{Liu \& Liu}{2010}]{Liu_SPH_kernel}
Liu M.,  Liu G.,  2010, \mn@doi [Archives of Computational Methods in Engineering] {10.1007/s11831-010-9040-7}, 17, 25

\bibitem[\protect\citeauthoryear{{Madau}, {Haardt}  \& {Dotti}}{{Madau} et~al.}{2014}]{Madau2014}
{Madau} P.,  {Haardt} F.,   {Dotti} M.,  2014, \mn@doi [\apjl] {10.1088/2041-8205/784/2/L38}, \href {https://ui.adsabs.harvard.edu/abs/2014ApJ...784L..38M} {784, L38}

\bibitem[\protect\citeauthoryear{{Matsuoka} et~al.,}{{Matsuoka} et~al.}{2018}]{Matsuoka2018_qso_LF}
{Matsuoka} Y.,  et~al., 2018, \mn@doi [\apj] {10.3847/1538-4357/aaee7a}, \href {https://ui.adsabs.harvard.edu/abs/2018ApJ...869..150M} {869, 150}

\bibitem[\protect\citeauthoryear{{Matsuoka} et~al.,}{{Matsuoka} et~al.}{2019}]{Matsuoka2019_highz_qso}
{Matsuoka} Y.,  et~al., 2019, \mn@doi [\apj] {10.3847/1538-4357/ab3c60}, \href {https://ui.adsabs.harvard.edu/abs/2019ApJ...883..183M} {883, 183}

\bibitem[\protect\citeauthoryear{{Mazzucchelli} et~al.,}{{Mazzucchelli} et~al.}{2017}]{Mazzucchelli2017}
{Mazzucchelli} C.,  et~al., 2017, \mn@doi [\apj] {10.3847/1538-4357/aa9185}, \href {https://ui.adsabs.harvard.edu/abs/2017ApJ...849...91M} {849, 91}

\bibitem[\protect\citeauthoryear{{Meiksin}, {Tittley}  \& {Brown}}{{Meiksin} et~al.}{2010}]{Meiksin2010_thermal_effect}
{Meiksin} A.,  {Tittley} E.~R.,   {Brown} C.~K.,  2010, \mn@doi [\mnras] {10.1111/j.1365-2966.2009.15667.x}, \href {https://ui.adsabs.harvard.edu/abs/2010MNRAS.401...77M} {401, 77}

\bibitem[\protect\citeauthoryear{{Miralda-Escud{\'e}} \& {Rees}}{{Miralda-Escud{\'e}} \& {Rees}}{1998}]{Miralda-Escude1998}
{Miralda-Escud{\'e}} J.,  {Rees} M.~J.,  1998, \mn@doi [\apj] {10.1086/305458}, \href {https://ui.adsabs.harvard.edu/abs/1998ApJ...497...21M} {497, 21}

\bibitem[\protect\citeauthoryear{{Morey}, {Eilers}, {Davies}, {Hennawi}  \& {Simcoe}}{{Morey} et~al.}{2021}]{Morey2021_qso_tq_nz}
{Morey} K.~A.,  {Eilers} A.-C.,  {Davies} F.~B.,  {Hennawi} J.~F.,   {Simcoe} R.~A.,  2021, \mn@doi [\apj] {10.3847/1538-4357/ac1c70}, \href {https://ui.adsabs.harvard.edu/abs/2021ApJ...921...88M} {921, 88}

\bibitem[\protect\citeauthoryear{{Mortlock} et~al.,}{{Mortlock} et~al.}{2011}]{Mortlock2011_qso_70}
{Mortlock} D.~J.,  et~al., 2011, \mn@doi [\nat] {10.1038/nature10159}, \href {https://ui.adsabs.harvard.edu/abs/2011Natur.474..616M} {474, 616}

\bibitem[\protect\citeauthoryear{{Nelson} et~al.,}{{Nelson} et~al.}{2015}]{Nelson2015_illustris}
{Nelson} D.,  et~al., 2015, \mn@doi [Astronomy and Computing] {10.1016/j.ascom.2015.09.003}, \href {https://ui.adsabs.harvard.edu/abs/2015A&C....13...12N} {13, 12}

\bibitem[\protect\citeauthoryear{Ni, Di Matteo  \& Feng}{Ni et~al.}{2021}]{Ni2021_gaussianCR}
Ni Y.,  Di Matteo T.,   Feng Y.,  2021, \mn@doi [MNRAS] {10.1093/mnras/stab3162}, 509, 3043

\bibitem[\protect\citeauthoryear{{Ni} et~al.,}{{Ni} et~al.}{2022}]{Ni2022_astrid}
{Ni} Y.,  et~al., 2022, \mn@doi [\mnras] {10.1093/mnras/stac351}, \href {https://ui.adsabs.harvard.edu/abs/2022MNRAS.513..670N} {513, 670}

\bibitem[\protect\citeauthoryear{{Novak}, {Ostriker}  \& {Ciotti}}{{Novak} et~al.}{2011}]{Novak2011_lc_vary}
{Novak} G.~S.,  {Ostriker} J.~P.,   {Ciotti} L.,  2011, \mn@doi [\apj] {10.1088/0004-637X/737/1/26}, \href {https://ui.adsabs.harvard.edu/abs/2011ApJ...737...26N} {737, 26}

\bibitem[\protect\citeauthoryear{{Okamoto}, {Frenk}, {Jenkins}  \& {Theuns}}{{Okamoto} et~al.}{2010}]{Okamoto2010_SNII_feedback}
{Okamoto} T.,  {Frenk} C.~S.,  {Jenkins} A.,   {Theuns} T.,  2010, \mn@doi [\mnras] {10.1111/j.1365-2966.2010.16690.x}, \href {https://ui.adsabs.harvard.edu/abs/2010MNRAS.406..208O} {406, 208}

\bibitem[\protect\citeauthoryear{{Oppenheimer}, {Segers}, {Schaye}, {Richings}  \& {Crain}}{{Oppenheimer} et~al.}{2018}]{Oppenheimer2018_qso_lc}
{Oppenheimer} B.~D.,  {Segers} M.,  {Schaye} J.,  {Richings} A.~J.,   {Crain} R.~A.,  2018, \mn@doi [\mnras] {10.1093/mnras/stx2967}, \href {https://ui.adsabs.harvard.edu/abs/2018MNRAS.474.4740O} {474, 4740}

\bibitem[\protect\citeauthoryear{{Read}, {Hayfield}  \& {Agertz}}{{Read} et~al.}{2010}]{Read2010_pSPH}
{Read} J.~I.,  {Hayfield} T.,   {Agertz} O.,  2010, \mn@doi [\mnras] {10.1111/j.1365-2966.2010.16577.x}, \href {https://ui.adsabs.harvard.edu/abs/2010MNRAS.405.1513R} {405, 1513}

\bibitem[\protect\citeauthoryear{{Reed} et~al.,}{{Reed} et~al.}{2017}]{Reed2017_highz_qso}
{Reed} S.~L.,  et~al., 2017, \mn@doi [\mnras] {10.1093/mnras/stx728}, \href {https://ui.adsabs.harvard.edu/abs/2017MNRAS.468.4702R} {468, 4702}

\bibitem[\protect\citeauthoryear{{Regan}, {Downes}, {Volonteri}, {Beckmann}, {Lupi}, {Trebitsch}  \& {Dubois}}{{Regan} et~al.}{2019}]{Regan2019}
{Regan} J.~A.,  {Downes} T.~P.,  {Volonteri} M.,  {Beckmann} R.,  {Lupi} A.,  {Trebitsch} M.,   {Dubois} Y.,  2019, \mn@doi [\mnras] {10.1093/mnras/stz1045}, \href {https://ui.adsabs.harvard.edu/abs/2019MNRAS.486.3892R} {486, 3892}

\bibitem[\protect\citeauthoryear{{Rudd}, {Zentner}  \& {Kravtsov}}{{Rudd} et~al.}{2008}]{rudd2008}
{Rudd} D.~H.,  {Zentner} A.~R.,   {Kravtsov} A.~V.,  2008, \mn@doi [\apj] {10.1086/523836}, \href {https://ui.adsabs.harvard.edu/abs/2008ApJ...672...19R} {672, 19}

\bibitem[\protect\citeauthoryear{{Rumbaugh} et~al.,}{{Rumbaugh} et~al.}{2018}]{Rumbaugh2018_qso_vary}
{Rumbaugh} N.,  et~al., 2018, \mn@doi [\apj] {10.3847/1538-4357/aaa9b6}, \href {https://ui.adsabs.harvard.edu/abs/2018ApJ...854..160R} {854, 160}

\bibitem[\protect\citeauthoryear{{Satyavolu}, {Kulkarni}, {Keating}  \& {Haehnelt}}{{Satyavolu} et~al.}{2023}]{Satyavolu2023_qso_growth}
{Satyavolu} S.,  {Kulkarni} G.,  {Keating} L.~C.,   {Haehnelt} M.~G.,  2023, \mn@doi [\mnras] {10.1093/mnras/stad729}, \href {https://ui.adsabs.harvard.edu/abs/2023MNRAS.521.3108S} {521, 3108}

\bibitem[\protect\citeauthoryear{{Schawinski}, {Koss}, {Berney}  \& {Sartori}}{{Schawinski} et~al.}{2015}]{Schawinski2015_qso_lc}
{Schawinski} K.,  {Koss} M.,  {Berney} S.,   {Sartori} L.~F.,  2015, \mn@doi [\mnras] {10.1093/mnras/stv1136}, \href {https://ui.adsabs.harvard.edu/abs/2015MNRAS.451.2517S} {451, 2517}

\bibitem[\protect\citeauthoryear{{Schleicher}, {Palla}, {Ferrara}, {Galli}  \& {Latif}}{{Schleicher} et~al.}{2013}]{Schleicher2013_dc_bh_seed}
{Schleicher} D. R.~G.,  {Palla} F.,  {Ferrara} A.,  {Galli} D.,   {Latif} M.,  2013, \mn@doi [\aap] {10.1051/0004-6361/201321949}, \href {https://ui.adsabs.harvard.edu/abs/2013A&A...558A..59S} {558, A59}

\bibitem[\protect\citeauthoryear{{Shakura} \& {Sunyaev}}{{Shakura} \& {Sunyaev}}{1973}]{Shakura1973_BH}
{Shakura} N.~I.,  {Sunyaev} R.~A.,  1973, \aap, \href {https://ui.adsabs.harvard.edu/abs/1973A&A....24..337S} {24, 337}

\bibitem[\protect\citeauthoryear{{Shen}}{{Shen}}{2021}]{Shen2021_qso_lc}
{Shen} Y.,  2021, \mn@doi [\apj] {10.3847/1538-4357/ac1ce4}, \href {https://ui.adsabs.harvard.edu/abs/2021ApJ...921...70S} {921, 70}

\bibitem[\protect\citeauthoryear{{Smith}, {Bromm}  \& {Loeb}}{{Smith} et~al.}{2017}]{Smith2017}
{Smith} A.,  {Bromm} V.,   {Loeb} A.,  2017, \mn@doi [Astronomy and Geophysics] {10.1093/astrogeo/atx099}, \href {https://ui.adsabs.harvard.edu/abs/2017A&G....58c3.22S} {58, 3.22}

\bibitem[\protect\citeauthoryear{{Springel} \& {Hernquist}}{{Springel} \& {Hernquist}}{2003}]{Springel2003_SF_model}
{Springel} V.,  {Hernquist} L.,  2003, \mn@doi [\mnras] {10.1046/j.1365-8711.2003.06206.x}, \href {https://ui.adsabs.harvard.edu/abs/2003MNRAS.339..289S} {339, 289}

\bibitem[\protect\citeauthoryear{{Springel}, {Di Matteo}  \& {Hernquist}}{{Springel} et~al.}{2005}]{Springel2005_bh_model}
{Springel} V.,  {Di Matteo} T.,   {Hernquist} L.,  2005, \mn@doi [\mnras] {10.1111/j.1365-2966.2005.09238.x}, \href {https://ui.adsabs.harvard.edu/abs/2005MNRAS.361..776S} {361, 776}

\bibitem[\protect\citeauthoryear{{Stevans}, {Shull}, {Danforth}  \& {Tilton}}{{Stevans} et~al.}{2014}]{stevans_2014_qso_fesc}
{Stevans} M.~L.,  {Shull} J.~M.,  {Danforth} C.~W.,   {Tilton} E.~M.,  2014, \mn@doi [\apj] {10.1088/0004-637X/794/1/75}, \href {https://ui.adsabs.harvard.edu/abs/2014ApJ...794...75S} {794, 75}

\bibitem[\protect\citeauthoryear{{Tepper-Garc{\'\i}a}}{{Tepper-Garc{\'\i}a}}{2006}]{Tepper-Garcia2006_Voigt_profile}
{Tepper-Garc{\'\i}a} T.,  2006, \mn@doi [\mnras] {10.1111/j.1365-2966.2006.10450.x}, \href {https://ui.adsabs.harvard.edu/abs/2006MNRAS.369.2025T} {369, 2025}

\bibitem[\protect\citeauthoryear{{Vogelsberger}, {Genel}, {Sijacki}, {Torrey}, {Springel}  \& {Hernquist}}{{Vogelsberger} et~al.}{2013}]{Vogelsberger2013_SF_model_modification}
{Vogelsberger} M.,  {Genel} S.,  {Sijacki} D.,  {Torrey} P.,  {Springel} V.,   {Hernquist} L.,  2013, \mn@doi [\mnras] {10.1093/mnras/stt1789}, \href {https://ui.adsabs.harvard.edu/abs/2013MNRAS.436.3031V} {436, 3031}

\bibitem[\protect\citeauthoryear{{Vogelsberger} et~al.,}{{Vogelsberger} et~al.}{2014}]{Vogelsberger2014_gas_metal_cool}
{Vogelsberger} M.,  et~al., 2014, \mn@doi [\mnras] {10.1093/mnras/stu1536}, \href {https://ui.adsabs.harvard.edu/abs/2014MNRAS.444.1518V} {444, 1518}

\bibitem[\protect\citeauthoryear{{Volonteri}}{{Volonteri}}{2010}]{Volonteri2010_SMBH_model}
{Volonteri} M.,  2010, \mn@doi [\aapr] {10.1007/s00159-010-0029-x}, \href {https://ui.adsabs.harvard.edu/abs/2010A&ARv..18..279V} {18, 279}

\bibitem[\protect\citeauthoryear{{Volonteri}, {Silk}  \& {Dubus}}{{Volonteri} et~al.}{2015}]{Volonteri2015}
{Volonteri} M.,  {Silk} J.,   {Dubus} G.,  2015, \mn@doi [\apj] {10.1088/0004-637X/804/2/148}, \href {https://ui.adsabs.harvard.edu/abs/2015ApJ...804..148V} {804, 148}

\bibitem[\protect\citeauthoryear{{Wang} et~al.,}{{Wang} et~al.}{2019}]{Wang2019_highz_qso}
{Wang} F.,  et~al., 2019, \mn@doi [\apj] {10.3847/1538-4357/ab2be5}, \href {https://ui.adsabs.harvard.edu/abs/2019ApJ...884...30W} {884, 30}

\bibitem[\protect\citeauthoryear{{Wang} et~al.,}{{Wang} et~al.}{2020}]{Wang2020_dampingwing}
{Wang} F.,  et~al., 2020, \mn@doi [\apj] {10.3847/1538-4357/ab8c45}, \href {https://ui.adsabs.harvard.edu/abs/2020ApJ...896...23W} {896, 23}

\bibitem[\protect\citeauthoryear{{Weinberger} et~al.,}{{Weinberger} et~al.}{2018}]{Weinberger2018}
{Weinberger} R.,  et~al., 2018, \mn@doi [\mnras] {10.1093/mnras/sty1733}, \href {https://ui.adsabs.harvard.edu/abs/2018MNRAS.479.4056W} {479, 4056}

\bibitem[\protect\citeauthoryear{{Worseck} et~al.,}{{Worseck} et~al.}{2014}]{Worseck2014_qso_fesc}
{Worseck} G.,  et~al., 2014, \mn@doi [\mnras] {10.1093/mnras/stu1827}, \href {https://ui.adsabs.harvard.edu/abs/2014MNRAS.445.1745W} {445, 1745}

\bibitem[\protect\citeauthoryear{{Worseck}, {Khrykin}, {Hennawi}, {Prochaska}  \& {Farina}}{{Worseck} et~al.}{2021}]{Worseck2021_qso_dating_heii}
{Worseck} G.,  {Khrykin} I.~S.,  {Hennawi} J.~F.,  {Prochaska} J.~X.,   {Farina} E.~P.,  2021, \mn@doi [\mnras] {10.1093/mnras/stab1685}, \href {https://ui.adsabs.harvard.edu/abs/2021MNRAS.505.5084W} {505, 5084}

\bibitem[\protect\citeauthoryear{{Wyithe}, {Loeb}  \& {Carilli}}{{Wyithe} et~al.}{2005}]{Wyithe2005}
{Wyithe} J. S.~B.,  {Loeb} A.,   {Carilli} C.,  2005, \mn@doi [\apj] {10.1086/430874}, \href {https://ui.adsabs.harvard.edu/abs/2005ApJ...628..575W} {628, 575}

\bibitem[\protect\citeauthoryear{{Yang} et~al.,}{{Yang} et~al.}{2020}]{Yang2020_dampingwing}
{Yang} J.,  et~al., 2020, \mn@doi [\apjl] {10.3847/2041-8213/ab9c26}, \href {https://ui.adsabs.harvard.edu/abs/2020ApJ...897L..14Y} {897, L14}

\bibitem[\protect\citeauthoryear{{{\v{S}}oltinsk{\'y}}, {Bolton}, {Molaro}, {Hatch}, {Haehnelt}, {Keating}, {Kulkarni}  \& {Puchwein}}{{{\v{S}}oltinsk{\'y}} et~al.}{2023}]{Soltinsky2023_qso_lifetime}
{{\v{S}}oltinsk{\'y}} T.,  {Bolton} J.~S.,  {Molaro} M.,  {Hatch} N.,  {Haehnelt} M.~G.,  {Keating} L.~C.,  {Kulkarni} G.,   {Puchwein} E.,  2023, \mn@doi [\mnras] {10.1093/mnras/stac3710}, \href {https://ui.adsabs.harvard.edu/abs/2023MNRAS.519.3027S} {519, 3027}

\bibitem[\protect\citeauthoryear{{van de Weygaert} \& {Bertschinger}}{{van de Weygaert} \& {Bertschinger}}{1996}]{vandeWeygaert1996_CR_para}
{van de Weygaert} R.,  {Bertschinger} E.,  1996, \mn@doi [\mnras] {10.1093/mnras/281.1.84}, \href {https://ui.adsabs.harvard.edu/abs/1996MNRAS.281...84V} {281, 84}

\makeatother
\end{thebibliography}




\appendix

\section{$R_{\mathrm{p}}$ for  the lightbulb model}
\label{section: results: lightbulb compare}

\begin{figure}
	\includegraphics[width=\columnwidth]{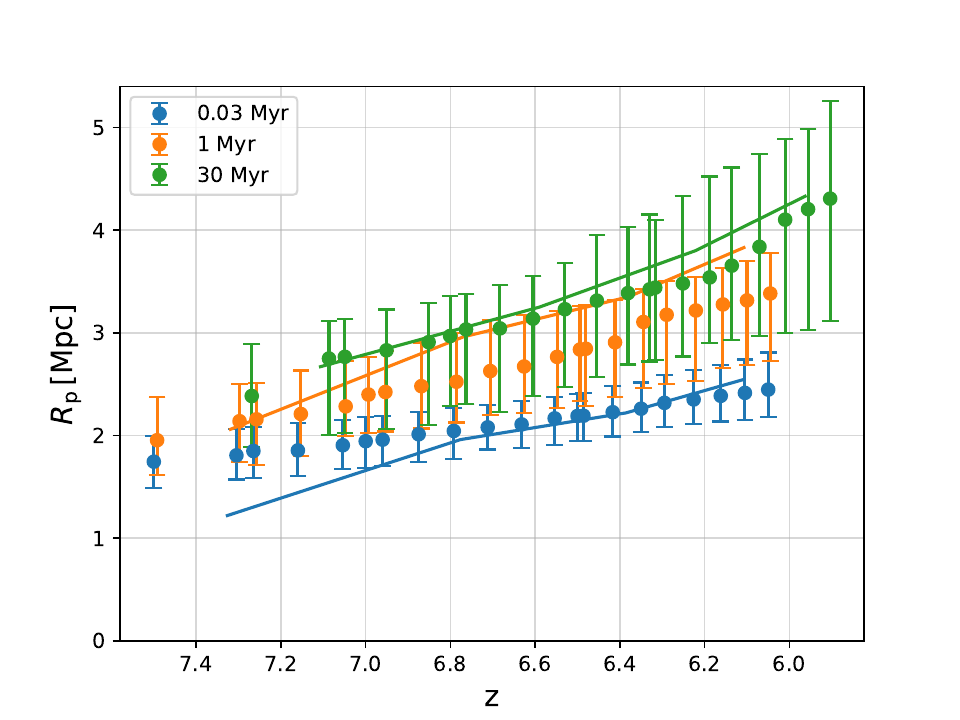}
    \caption{Based on a lightbulb model fixed at  $\dot{N}=1\times 10^{57}\ \mathrm{s}^{-1}$, the evolution of the proximity zone size $R_{\mathrm{p}}$ as a function of the quasar turn-on redshift for $t_{\mathrm{evol}}=0.03\ \mathrm{Myr}$ (blue), $1\ \mathrm{Myr}$ (orange), and $30\ \mathrm{Myr}$ (green). The dots are results from the CR simulation, whose errorbars represent the 16-84th scatter, and the solid curves are the median $R_{\mathrm{p}}$ modeled by \protect\cite{Chen2021_qso_prixmity_zone} with the CROC simulation. 
}
    \label{fig:lightbulb_test}
\end{figure}

In order to compare to previous work, we simulate proximity zone sizes for a lightbulb model with $\dot{N}=10^{57}\ \mathrm{s}^{-1}$ based on the CR simulation used in this paper.  For each line of sight, we turn on the quasars at each snapshot redshift $z_{\mathrm{snap}}$ and evolve the radiation for $30\ \mathrm{Myr}$. In Fig.~\ref{fig:lightbulb_test}, we show the measured $R_{\mathrm{p}}$ distribution after $t_{\mathrm{evol}}=0.03\ \mathrm{Myr}$ (blue), $t_{\mathrm{evol}}=1\ \mathrm{Myr}$ (orange), and $t_{\mathrm{evol}}=30\ \mathrm{Myr}$ (green) from the CR simulation using the dots, with the errorbar indicating the 16-84th percentiles of scatter. We also portray the median $R_{\mathrm{p}}$ modeled by \cite{Chen2021_qso_prixmity_zone} with the solid curves. 
 
Fig.\ref{fig:lightbulb_test} demonstrates that our simulation generates similar $R_{\mathrm{p}}$ to \cite{Chen2021_qso_prixmity_zone} at $t_{\mathrm{evol}}=0.03\ \mathrm{Myr}$ and $t_{\mathrm{evol}} = 30 \ \mathrm{Myr}$, while predicting smaller proximity zones at $t_{\mathrm{evol}} = 1 \ \mathrm{Myr}$. 
This difference in results probably stems from the more massive quasar host halo 
($\sim 10^{13}\ M_{\odot}$ rather than $\gtrsim 1.5 \times 10^{11}\ M_{\odot}$) we considered in our simulation, as shown in Fig.~\ref{fig:los_PDF}. 
With a given ionization fraction, the proximity zones in the high-density environment grow slower while reaching a similar maximum $R_{\mathrm{p}}$ with those in the low-density environment (see Fig. 7 in \cite{Keating2015}). 
At high redshift $z > 7$, our results seem to evolve faster than those produced by \cite{Chen2021_qso_prixmity_zone} at $t_{\mathrm{evol}} = 0.03$ Myr. This is because the CR simulation has completed reionization at $z=8$ (see Section~\ref{section:simu_phy}), while  CROC still has a large volume of neutral hydrogen in the IGM around $z=7.3$ ($x_{\mathrm{\hi}}=0.13$).

\section{Influence of background evolution}
\label{section: error_ana}

In this section, we explore the uncertainty in the simulated $R_{\mathrm{p}}$ caused by the stitching procedure applied in Section~\ref{section: results: proxmity zone evolution }. 
When we calculate the evolution of $R_{\mathrm{p}}$,
we only update the properties of the IGM (including the density and velocity field) at $z_{\mathrm{snap}}$, which breaks the continuous evolution of IGM
and might make our results differ from reality.

To estimate the uncertainty caused by this method of background updating, we use the semi-analytical model proposed in \cite{Davies2020_NR_RT_model} and calculate the largest change in $R_{\mathrm{p}}$ when the background is updated, which roughly equals to the largest error caused by the stitching procedure if we assume the simulated $R_{\mathrm{p}}$ at newly-updated $z_{\mathrm{snap}}$ is accurate.
Taking into account that the IGM environments adopted in \cite{Davies2020_NR_RT_model} are different from ours (such as the density, ionization fraction), we first re-scale the semi-analytical model for our simulation: 
\begin{equation}
\begin{aligned}    
R_{\mathrm{p}} &= 3.1544\left(\frac{\Gamma_{\mathrm{bkg}}^{\mathrm{\hi}}}{2.455\times 10^{-13}\,\text{s}^{-1}}\,\right)^{-1/2}
\left[\left(\frac{\tau_{\mathrm{bkg}}}{2.3}\right)^{1/\alpha}-1\right]^{-1/2}\\ 
&\times\left(\frac{\dot{N}}{1.3197\times10^{56}\,\text{s}^{-1}}\right)^{1/2}\left(\frac{1+z}{7.5}\right)^{-3/2}\ \mathrm{Mpc},
\end{aligned}\label{equ: semi-ana_davies}
\end{equation}
where $\tau_{\mathrm{bkg}}=5.678\left(\Gamma^{\mathrm{\hi}}_{\mathrm{bkg}}/\left[2.5\times10^{-13}\,\mathrm{s}^{-1}\right]\right)^{-\alpha}$ and $\alpha = 0.5486$.
In this model, all of the influences of the background are encoded in $\Gamma_{\mathrm{bkg}}^{\mathrm{\hi}}$, which is 
the collisional ionization rate of hydrogen caused by the background radiation. We calculate the value of $\Gamma_{\mathrm{bkg}}^{\mathrm{\hi}}$ in the same way as \cite{Chen2021_qso_prixmity_zone} (see their equation 3), which is determined by the density, temperature and the ionization fraction of the IGM without the quasar radiation. 
The largest time gap between two consecutive snapshots is $25\ \mathrm{Myr}$, which exists between $z_{\mathrm{snap}}=7.500$ and $z_{\mathrm{snap}}=7.305$, corresponding to a change in  $\Gamma_{\mathrm{bkg}}^{\mathrm{\hi}}$ of $\lesssim 2 \times 10^{-14}\ \mathrm{s}^{-1}$. 
Based on equation~\ref{equ: semi-ana_davies},
such change in $\Gamma_{\mathrm{bkg}}^{\mathrm{\hi}}$ results in an uncertainty of $\Delta R_{\mathrm{p}}\sim 0.05\ \mathrm{Mpc}$. 
Compared with the $R_{\mathrm{p}}$ scatter stemming from the density field fluctuations and the light curve variation, both of which are $\sim 0.3$ Mpc as mentioned in Section~\ref{section: results: proxmity zone evolution }, such uncertainty is insignificant.   
Hence, the background evolution between two adjacent snapshots has negligible effects on our simulation.


\section{The Influence of Variation Amplitudes in Quasar Light Curves}
\label{app: var_mag}


Here we briefly discuss how the variation amplitude in the light curve influences the amount of discrepancy between the variable light curve and the lightbulb models. 
We build toy models for three light curves 
varying around $M_{\mathrm{UV}}=-25.5$ with different variation amplitudes: $\Delta M_{\mathrm{UV}} =\pm 1$, $\Delta M_{\mathrm{UV}} = \pm 2$, and $\Delta M_{\mathrm{UV}} =\pm 3$, respectively.
These light curves have zigzag shapes, whose magnitudes change linearly with time from $-25.5 + |\Delta M_{\mathrm{UV}}|$ to $-25.5 - |\Delta M_{\mathrm{UV}}|$, and then drop linearly to $-25.5 + |\Delta M_{\mathrm{UV}}|$ to complete a period.
They all have a fluctuation period of $0.1\ \mathrm{Myr}$ and we evolve them for $0.5$ Myr.
We calculate the $R_{\mathrm{p}}$ generated by these light curves on the lines of sight from $z_{\mathrm{snap}}=6.162$ snapshot with a time resolution of $2\times 10^{3}\ \mathrm{yr}$.
By comparing the resultant $R_{\mathrm{p}}$ distributions, we are allowed to see the influence of variation amplitudes.

\begin{figure}
	\includegraphics[width=\columnwidth]{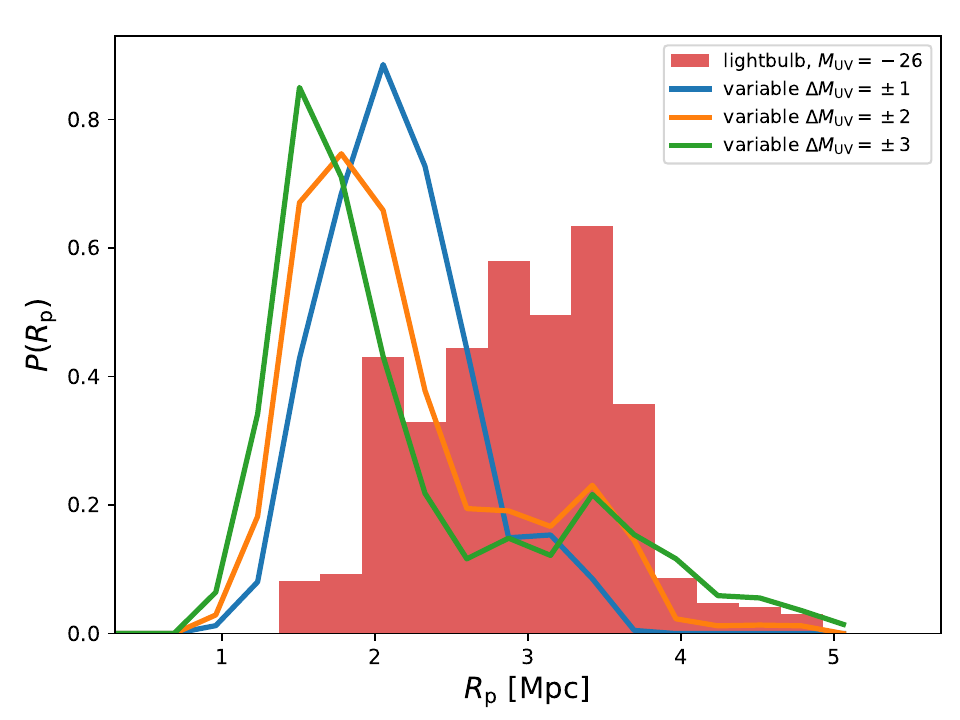}
    \caption{
    The $R_{\mathrm{p}}$ distribution generated by the variable light curves with $-26.5 < M_{\mathrm{UV}} <  -25.5$ (solid curves) and by a lightbulb model fixed at $M_{\mathrm{UV}}=-26$ (red shade area). 
    The three variable light curves have a fixed fluctuation period of $10^{5}$ yr, and vary around $M_{\mathrm{UV}}=-25.5$ with the variation amplitudes of $\Delta M_{\mathrm{UV}} = \pm 1$ (blue), $\pm 2$ (orange), $\pm 3$ (green), respectively. 
    All the light curves evolve $0.5\ \mathrm{Myr}$.
    The $R_{\mathrm{p}}$ are computed on the $z_{\mathrm{snap}}=6.162$ snapshot and are sampled with a time resolution of $2\times 10^{3}\ \mathrm{yr}$.
}
    \label{fig:Robs_hist_1D}
\end{figure}

In Fig.~\ref{fig:Robs_hist_1D}, we depict the one-dimensional histogram for the $R_{\mathrm{p}}$ sampling from the variable light curves within $-26.5<M_{\mathrm{UV}}<-25.5$ (solid curves) as well as those produced by a lightbulb model fixed at $M_{\mathrm{UV}}=-26$ (red shaded area).
The mean $R_{\mathrm{p}}$ for the lightbulb model is $2.93$ pMpc. 
While for the variable light curves with $\Delta M_{\mathrm{UV}}=\pm 1,\ \pm 2,\  \pm 3$, the mean $R_{\mathrm{p}}$ are $2.15$ pMpc, $2.19$ pMpc, and $2.32$ pMpc, respectively. 
All the variable light curves generate an $R_{\mathrm{p}}$ distribution peaking at smaller values. Furthermore, larger variation magnitudes result in a slightly larger shift from the lightbulb model. 

\bsp	
\label{lastpage}
\end{document}